\documentclass{article}
\usepackage{amsmath}
\usepackage{amssymb}
\usepackage{amsfonts}
\usepackage{mathrsfs}
\usepackage{theorem}
\usepackage{graphicx}
\usepackage{caption}
\usepackage{subcaption}
\usepackage{epstopdf}
\usepackage{geometry}
\usepackage{hyperref}

\newtheorem{theorem}{\bf Theorem}[section]

\theorembodyfont{\rmfamily}

\newtheorem{remark}[theorem]{Remark}

\newcommand{\R}{\mbox{${\mathcal R}$}}

\begin{document}

\title{Numerical methods for simulating the motion of porous balls in simple 3D shear flows under creeping conditions}

\author{Aixia Guo,  Tsorng-Whay Pan\footnotemark[1]  \  and Jiwen He\\
{\it  \large Department of Mathematics, University of Houston, Houston, Texas 77204, USA} \\ \\
  Roland Glowinski\\
{\it \large Department of Mathematics, University of Houston, Houston, Texas 77204, USA}\\
{\it \large Department of Mathematics, Baptist University, Hong-Kong}}

\date{}
\maketitle

\begin{abstract}
In this article, two novel numerical methods have been developed for simulating fluid/porous particle 
interactions in three-dimensional (3D) Stokes flow. The Brinkman-Debye-Bueche model is adopted for the 
fluid flow inside the porous particle, being coupled with the Stokes equations for the fluid flow 
outside the particle. The rotating motion of a porous ball and  the interaction of two porous balls in bounded 
shear flows have been studied by these two new methods. The numerical results show
that the porous particle permeability   has a strong effect on the interaction of two porous balls.
\end{abstract}

\vskip 4.5mm
\noindent{{\bf Keywords } Porous particles;  Brinkman-Debye-Bueche model;  Stokes flow,  Particle suspension.

\noindent{{\bf Classification:} AMS: 65M60}

\baselineskip 14pt

\setlength{\parindent}{1.5em}

\section{Introduction}
The study of interactions of fluid flow with porous particles is of great interest. Porous materials are everywhere. 
The skin, bones and a lot of organs of human body are porous. Daily foods like bread, vegetables and meats and  building materials such 
as bricks, concrete and limestone are also porous. In medicine and biomedical engineering, biological membranes and filters provide other examples of porous materials. 
In electrochemical processes, permeable diaphragms and electrodes are further examples of porous materials. There are a lot of practical cases about 
flow around and through porous materials in industrial processes: A typical example concerns the production of paper from a pulp suspension; indeed, the flow of water 
around and through a pulp floc is a flow around and within a porous medium.  
In \cite{2, 3, 4, 5, 6, 7, 8, 9, 10} there are studies of the porous contexts in agglomeration, chromatography, drug diliverly and tissue engineering, pulp 
and paper manufacturing and waste-water treatment. Darcy's law \cite{11} is the most common model of porous media flow studies
and is a reliable formula as discussed in \cite{30}. However,   Darcy's law has some limitations: For example, velocity gradients cannot be dealt with due to 
shear like those that occur at the  boundary of the porous materials. Moreover, it is impossible for Darcy's law to match the velocity and shear stress continuously for the interior and 
exterior flows at the particle surface since the Darcy's law is a first order equation and Stokes equations, which govern the flow outside the porous particles,
is second order. In order to solve the limitation on porous surface, Brinkman \cite{12,13} and Debue \& Bueche \cite{14} included the velocity gradient, 
which is not present in Darcy's equation, and introduced the so-called Brinkman-Debye-Bueche (BDB) model. Since then, many  theoretical studies have been focused
on  flow past rigid porous bodies, including \cite{15, 17, 18, 19, 20, 21, 29} for  single particle flow, and \cite{22, 23, 24}  for   flow past two porous particles.
It is important to understand the rheology of the suspension of porous particles  in fluid flow, one major application
being that porous particles can be used as carriers for drug delivery as, e.g., in \cite{Tasciotti2008, Zhang2015}. 
In \cite{29}, Masoud {\it et al.} studied numerically the influence of the  permeability on the rotation rate of a porous ellipsoid in simple shear 
flows at the Stokes regime, based on the Brinkman-Debye-Bueche model. In \cite{Li2016}, Li {\it et al.} studied the effect of the fluid  inertia 
on the rotational behavior of a circular porous particle suspended in a two-dimensional 
simple shear flow.   In this article, the Brinkman-Debye-Bueche model is adopted for the 
fluid flow inside the porous particle and then coupled with the Stokes equations for the fluid flow 
outside the particle. We have developed two novel numerical methods for simulating fluid/porous particle 
interactions in three-dimensions under creeping flow conditions.  The rotation of a porous ball and the interaction of two porous balls in bounded 
shear flows have been studied. Numerical results show that, due to the permeability, two porous balls interact in bounded shear flows in a way
quite different from that of two non-porous and rigid balls.
The outline of this article is as follows:  In Section 2, we will provide the system of equations modeling the fluid/porous particle interaction.
Next, in Section 3 we will describe our numerical methods and present the results of  the numerical simulation of the motion of one and then 
two porous balls in a bounded shear flow.

\section{Governing equations}\label{sec.1}

\begin{figure}[ht!]
\begin{center}
\includegraphics[width=0.5\textwidth]{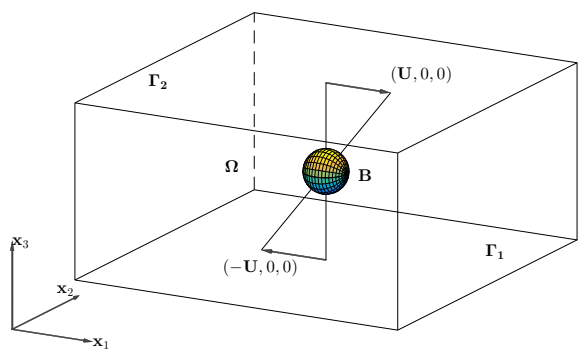}
\end{center}
\caption{An example of a shear flow region with a porous ball $B$.} \label{fig.1}
\end{figure}

Without loss of generality, let us consider a neutrally buoyant and permeable ball $B$ moving in a bounded shear flow shown in Fig. \ref{fig.1}
where $\Omega \subset \R^3$ is a rectangular parallelepiped  filled with a Newtonian viscous 
incompressible fluid. The flow governing equations outside the particle $B(t)$ are the following Stokes equations
\begin{eqnarray}
&&\nabla \cdot \boldsymbol \sigma \equiv \nu \nabla^2 \mathbf u -\nabla p = {\bf 0} \text{ in } \Omega \setminus \overline{B(t)}, \, \, t \in (0,T), \label{eqn:1}\\
&&\nabla \cdot \mathbf{u} =0 \text{ in } \Omega \setminus \overline{B(t)}, \, \, t \in (0,T), \label{eqn:2}\\
&& {\bf u}= {\bf g}_0 \ \ {\rm on} \ \ \Gamma \times (0,T), \ {\rm with} \ \displaystyle\int_\Gamma {\bf g}_0 \cdot {\bf n} \,d \Gamma =0, \label{eqn:3}
\end{eqnarray}
where $\bf u$ is the flow velocity,  $p$ is 
the pressure,    $\nu$ is the fluid viscosity coefficient,  
$\Gamma$  is the union of the bottom boundary $\Gamma_1$ and top boundary   $\Gamma_2$ as in 
Fig. \ref{fig.1}, ${\bf n}$ is the unit normal vector pointing outward to the flow region, the boundary 
conditions being ${\bf g}_0=\{-U, 0, 0\}^t$ on $\Gamma_1$ and ${\bf g}_0=\{U, 0, 0\}^t$ on $\Gamma_2$ for a
bounded shear flow. We assume also that the flow is periodic in the $x_1$ and $x_2$ directions with 
the periods $L_1$ and $L_2$, respectively.

According to the BDB model (e.g., see \cite{29}), the fluid flow inside the porous particle is governed by the equations
\begin{eqnarray}
&&\nabla \cdot \boldsymbol \sigma \equiv \nu \nabla^2 \mathbf u -\nabla p = \nu k^{-1} (\mathbf u-\mathbf u_p)\ \text{ in } B(t), \, \, t \in (0,T), \label{eqn:4}\\
&&\nabla \cdot \mathbf u =0 \  \text{ in } B(t), \, \, t \in (0,T), \label{eqn:5}
\end{eqnarray}
where $k$ is the Darcy permeability and $\bf{u}_p$ is the velocity of the particle skeleton. The  BDB and Stokes equations are coupled through the 
continuity of velocity and traction forces at the particle surface. The aforementioned governing
equations for the fluid flow inside and outside of the porous particle have been considered in \cite{29} for investigating the rotational speed
of a porous particle of prolate ellipsoid shape in simple shear flows.

The motion of a neutrally buoyant ball satisfies the Euler-Newton's equations:
\begin{eqnarray}
&&\frac{d\mathbf G}{dt}=\mathbf V,\label{eqn:6}\\
&&M_p \frac{d\mathbf V}{dt}=\mathbf F_H,\label{eqn:7}\\
&&{\mathbf I}_p\frac{d\boldsymbol {\omega}}{dt}=\mathbf T_H,\label{eqn:8}\\
&&\mathbf G(0)=\mathbf G_0, \mathbf V(0)=\mathbf V_0, \boldsymbol \omega(0)=\boldsymbol \omega_0,\label{eqn:9}
\end{eqnarray}
where $\mathbf G$ is the mass center of the porous particle, $M_p$ is the particle mass, $\mathbf I_p$ is the inertia tensor, ${\mathbf V}$ 
is the velocity of the mass center and $\boldsymbol \omega$ is the particle angular velocity. The motion $\mathbf u_p$ of the porous particle skeleton  is given by 
\begin{equation}
\mathbf u_p(\mathbf x,t)=\mathbf V(t)+ \boldsymbol \omega(t)\times \overrightarrow{\mathbf{Gx}}, \forall {\mathbf x} \in \overline{B(t)},\, \, t \in (0,T).\label{eqn:10}
\end{equation}
For the hydrodynamical force and its resulting torque  acting on the  porous ball  in (\ref{eqn:7}) and (\ref{eqn:8}),  we have 
\begin{eqnarray}
&&\mathbf F_H =\int_{\partial B} \boldsymbol \sigma \mathbf n dS =\int_B \nabla \cdot \boldsymbol \sigma \, d \mathbf x= \nu k^{-1} \int_B(\mathbf u-\mathbf u_p) \, d \mathbf x,\label{eqn:11}\\
&&\mathbf T_H =\int_{B} \overrightarrow{\mathbf{Gx}} \times (\nabla \cdot  \boldsymbol \sigma) d\mathbf x= \nu  k^{-1}\int_B \overrightarrow{\mathbf{Gx}} \times(\mathbf u-\mathbf u_p) d\mathbf x,\label{eqn:12}
\end{eqnarray}
where $\mathbf n=\{n_1, n_2, n_3\}^t$ is the unit normal vector pointing 
outward to the flow region.

The variational formulation for a porous particle
freely moving in a Newtonian fluid under the creeping flow conditions is as follows:
\vskip 2ex
{\it For a.e. $t>0$, find ${\mathbf u}(t) \in (H^1(\Omega))^3, {\mathbf u}={\mathbf g}_0(t)$ on 
$\Gamma$, $p \in L^2_0 (\Omega), \mathbf{G(t), V(t)}, \boldsymbol{\omega(t)} \in \mathbb R^3$ such that}
\begin{eqnarray}
&&\nu \int_\Omega \nabla \mathbf u : \nabla \mathbf v\, d \mathbf x - \int_{\Omega} p \nabla \cdot \mathbf v\, d \mathbf x = -\frac{\nu}{k} \int_{B(t)} (\mathbf u-\mathbf u_p)\cdot \mathbf v\, d \mathbf x, \forall \mathbf v \in \mathbf W_0, \label{eqn:13}\\
&&\int_\Omega q \nabla \cdot \mathbf u \, d \mathbf x =0, \forall q \in L^2(\Omega), \label{eqn:14}\\
&&\frac{d\mathbf G}{dt}=\mathbf V, \label{eqn:15}\\
&&M_p \frac{d\mathbf V}{dt}=\mathbf F_H, \label{eqn:16}\\
&&{\mathbf I}_p\frac{d \boldsymbol {\omega}}{dt}=\mathbf T_H, \label{eqn:17} \\
&&\mathbf G(0)=\mathbf G_0, \mathbf V(0)=\mathbf V_0, \boldsymbol \omega(0)=\boldsymbol \omega_0, \label{eqn:18}
\end{eqnarray}
where in equations $(\ref{eqn:13})-(\ref{eqn:18})$, the function spaces are defined by
\begin{eqnarray*}
&&\mathbf W_0=\{\mathbf v \in (H^1(\Omega))^3,\ \mathbf v |_\Gamma =\mathbf 0, \ \mathbf v \text{ is periodic in the } x_1 \text{ and } x_2 \text{ directions }\\
&& \hskip 40pt \text{with periods $L_1$ and $L_2$, respectively}\},\\
&&L^2_0(\Omega) = \{q|q \in L^2(\Omega), \int_\Omega q \, d \mathbf x =0\},
\end{eqnarray*}
and
\begin{eqnarray*}
&&\mathbf u_p(\mathbf x,t)=\mathbf V(t)+ \boldsymbol \omega(t)\times \overrightarrow{\mathbf G(t)\mathbf x}, \forall {\mathbf x} \in B(t),\\
&&\mathbf F_H = \frac{\nu}{k} \int_{B(t)} (\mathbf u-\mathbf u_p)d\mathbf x,\\
&&\mathbf T_H = \frac{\nu}{k} \int_{B(t)} \overrightarrow{\mathbf G(t)\mathbf x} \times (\mathbf u-\mathbf u_p)d\mathbf x.
\end{eqnarray*}

We have also modified  formulation (\ref{eqn:13})-(\ref{eqn:18}) for simulating one ball freely rotating and suspended initially at the middle  between two 
moving walls in a bounded shear flow. To obtain the equilibrium rotational speed of such porous ball, we  consider the following time dependent Stokes flow problem:
\vskip 2ex

{\it For $a.e.\quad t>0$, find ${\mathbf u}(t) \in (H^1(\Omega))^3, {\mathbf u}={\mathbf g}_0(t)$ on $\Gamma$, 
$p \in L^2_0 (\Omega), \mathbf{G(t), V(t)}, \boldsymbol{\omega(t)} \in \mathbb R^3$ such that}
\begin{eqnarray}
&&\int_{\Omega} \frac{\partial \mathbf u}{\partial t} \cdot \mathbf v\, d \mathbf x + \nu \int_\Omega \nabla \mathbf u : \nabla \mathbf v\, d \mathbf x 
- \int_{\Omega} p \nabla \cdot \mathbf v\, d \mathbf x \nonumber \\
&& \hskip 100pt = -\frac{\nu}{k} \int_{B(t)} (\mathbf u-\mathbf u_p)\cdot \mathbf v\, d \mathbf x, \forall \mathbf v \in \mathbf W_0, \label{eqn:19}\\
&&\int_\Omega q (\nabla \cdot \mathbf u) \, d \mathbf x =0, \forall q \in L^2(\Omega), \label{eqn:20}\\
&&\frac{d\mathbf G}{dt}=\mathbf V(t), \label{eqn:21}\\
&&M_p \frac{d\mathbf V}{dt}=\mathbf F_H, \label{eqn:22}\\
&&\mathbf I_p \frac{d\boldsymbol {\omega}}{dt}=\mathbf T_H, \label{eqn:23} \\
&& \mathbf u(0)=\mathbf u_0 \quad (\text{with} \nabla \cdot \mathbf u_0=0) \text{ in } \Omega, \label{eqn:24}\\
&&\mathbf G(0)=\mathbf G_0, \mathbf V(0)=\mathbf V_0, \boldsymbol \omega(0)=\boldsymbol \omega_0.\label{eqn:25}
\end{eqnarray}
In equation (\ref{eqn:24}), $\mathbf u_0$ is the initial value of the velocity field $\mathbf u$.

\section{Numerical methods and results}

In this section, we like to present two novel numerical methodes for simulating the motion of porous balls in bounded shear flows based on the the coupled models 
discussed in Section \ref{sec.1}. For the case of a single porous ball freely rotating and suspended initially at the middle between two moving walls, its mass center remains 
there. Its rotating speed with respect to the vorticity direction is a constant. To catch such a motion, we have to reach the steady state solution of equations 
(\ref{eqn:19})-(\ref{eqn:25}); this can be done numerically. Similarly, to simulate the interaction of two porous balls in a bounded shear flow, we have to solve  
equations (\ref{eqn:13})-(\ref{eqn:18})  numerically. Therefore we have developed in this section two different algorithms for simulating these two kinds of 
fluid/porous particle interaction. In this section,  we assume that all dimensional quantities are in the CGS units. 

\subsection{Motion of a single porous ball}\label{sec3.1}

\subsubsection{Description of the first numerical method }

For the space discretization, we have used $P_1$-$iso$-$P_2$ and $P_1$ finite elements for the velocity field and pressure, respectively. More precisely, 
$h$ being the space discretization mesh size, we introduce a uniform tetrahedrization $\mathcal{T}_h$  and a twice coarser tetrahedrization 
$\mathcal{T}_{2h}$  of $\overline{\Omega}$. We approximate $(H^1(\Omega))^3, \mathbf W_0, L(\Omega)$ and $L^2_0(\Omega)$ by the following finite dimensional spaces
\begin{eqnarray*}
&&\mathbf W_h = \{\mathbf v_h| \mathbf v_h \in (C^0(\bar{\Omega}))^3, \mathbf v_h|_T \in (P_1)^3, \forall T \in \mathcal{T}_h, \mathbf v_h \text{ is 
periodic in the } x_1 \text{ and } x_2\\
&& \hskip 40pt \text{directions with periods $L_1$ and $L_2$, respectively}\},\\
&& \mathbf W_{0,h} = \{\mathbf v_h| \mathbf v_h \in \mathbf {\mathbf W}_h, \mathbf v_h = \mathbf 0|_{\Gamma}\},\\
&&L^2_h = \{q_h|q_h \in C^0(\bar\Omega), q_h|_T \in P_1, \forall T \in \mathcal{T}_{2h}\},\\
&&L^2_{0,h}=\{q_h|q_h \in L^2_h, \int_\Omega q_h \, d \mathbf x=0\},
\end{eqnarray*}
respectively;  above, $P_1$ is the space of polynomials in three variables of degree $\le 1.$

To simulate the motion of a single porous particle in a bounded shear flow, we look for a steady state of the rotation 
velocity field of the porous ball. To reach this steady state, we prefer working with  formulation (\ref{eqn:19})-(\ref{eqn:25}). 
Let $\Delta t$ be a time step and $t^n = n \Delta t$. The  algorithm for solving the fully discrete analogue of system   (\ref{eqn:19})-(\ref{eqn:25})  
reads as follows (after dropping some of subscripts $h$):

\vskip 2ex
$\mathbf u^{0}=\mathbf u_0$, $\mathbf G^0=\mathbf G_0$, $\mathbf V^{0}= \mathbf V_0$,  
$\boldsymbol{\omega}^0=\boldsymbol{\omega}_0$, $\mathbf u_p^0=\mathbf V^0+ \boldsymbol{\omega}^0 \times \overrightarrow{\mathbf G^0 \mathbf x}$ {\it are known}. 

$\textit{For } n \ge 0,  \mathbf u^n, \mathbf u^n_p, \mathbf V^n, \mathbf G^n \textit{ and } \boldsymbol{\omega^n} \textit{being known, we solve the following 
sub-problems:}$

$\textit{Step 1. We compute } \mathbf u^{n+1} \textit{ and } p^{n+1} \textit{ via the solution of}$

\begin{eqnarray}
&&\int_{\Omega} \frac{\mathbf u^{n+1}-\mathbf u^n}{\Delta t}\cdot \mathbf v\, d \mathbf x + \nu \int_\Omega \nabla \mathbf u^{n+1}: \nabla \mathbf v\, d \mathbf x - \int_\Omega p^{n+1} \nabla \cdot \mathbf v\, d \mathbf x \nonumber\\
&& \hskip 100pt  = -\frac{\nu}{k} \int_{B^n}(\mathbf u^n - \mathbf u^n_p)\cdot \mathbf v\, d \mathbf x, \forall \mathbf v \in \mathbf W_{0,h}, \label{eqn:26}\\ 
&&\int_{\Omega} q \nabla \cdot \mathbf u^{n+1} \, d \mathbf x =0, \forall q \in L^2_h;\label{eqn:27}\\
&& \mathbf u^{n+1} \in \mathbf W_h, \mathbf u^{n+1}=\mathbf g_0(t^{n+1}) \textit{ on } \Gamma, p^{n+1} \in L^2_{0,h}.\nonumber
\end{eqnarray}

$\textit{Step 2. Update the particle velocities and predict its mass center position:}$

\begin{eqnarray}
&& M_p \frac{\mathbf V^{n+1}-\mathbf V^n}{\Delta t}=\frac{\nu}{k} \int_{B^n}(\mathbf u^{n+1}-\mathbf u^n_p)d\mathbf x,\label{eqn:28}\\
&& \mathbf I_p \frac{\boldsymbol \omega ^{n+1}-\boldsymbol \omega^n}{\Delta t} =\frac{\nu}{k} \int_{B^n} \overrightarrow{\mathbf G^n\mathbf x} \times (\mathbf u^{n+1}-\mathbf u_p^n)\, d \mathbf x,\label{eqn:29}\\
&& \mathbf G^{n+1}=\mathbf G^n +\mathbf V^{n+1} \Delta t,\label{eqn:30}\\
\textit{and then set}\nonumber\\
&& \mathbf u^{n+1}_p = \mathbf V^{n+1} + \boldsymbol \omega^{n+1} \times \overrightarrow{\mathbf G^{n+1}x}, \forall \mathbf x \in B^{n+1}\label{eqn:31},
\end{eqnarray}
where $B^{n+1}$ is the new solid volume occupied by the porous ball centered at $\mathbf G^{n+1}$.

\begin{remark}
In equation (\ref{eqn:26}), instead of having $\mathbf u^{n+1}$ in the right-hand side of the equation, we have used $\mathbf u^n$ there so that a 
fast solver can be applied to solve the linear systems at each iteration of an Uzawa/preconditioned 
conjugate gradient algorithm operating in the space $L^2_{0h}$ as discussed in, e.g., \cite{glowinski2003}.   For the cases of a single porous ball 
suspended in a bounded shear flow, we want to obtain an equilibrium state, which is a spherical porous particle rotating at a constant speed at 
the middle between two moving planes. The stopping criterion for the steady state is set to be
\begin{equation*}
\frac{1}{\Delta t} \|\mathbf u^{n+1}-\mathbf u^n\| < CRIT,
\end{equation*}
where {\it CRIT} is the tolerance. The numerical results obtained in the Section \ref{sec3.1.2} have been obtaned with $CRIT=10^{-5}$.
\end{remark}

\begin{remark} When computing $\int_{B^n}(\mathbf u^n - \mathbf u^n_p) \cdot \mathbf v\, d \mathbf x$ in equation (\ref{eqn:26}), we have extended the integration to the entire 
computational domain $\Omega$ as follows
\begin{equation*}
\int_{B^n}(\mathbf u^n - \mathbf u^n_p) \cdot \mathbf v\, d \mathbf x=\int_{\Omega} {\chi}_{B^n}(\mathbf u^n - \mathbf u^n_p) \cdot \mathbf v\, d \mathbf x,
\end{equation*}
where  $\chi_{B^n}$ is the characteristic function of the solid volume
occupied by the ball centered at $\mathbf G^n$, i.e.,
\begin{equation*}
\chi_{B^n}(\mathbf x)=
\begin{cases}
1 \quad \text{ if } \mathbf x \in B^n,\\
0 \quad \text{ if } \mathbf x \notin B^n.
\end{cases}
\end{equation*}
Then we have applied the trapezoidal rule on each tetrahedron in the tetrahedrization $\mathcal{T}_h$. To obtain the hydrodynamical forces
and their resulting torques in equations  (\ref{eqn:28}) and (\ref{eqn:29}), we have applied similar approaches.
\end{remark}
\begin{remark}
For a rigid ball of radius $a$, the mass and inertia tensor are
\begin{eqnarray*}
&&M_P = \frac{4}{3} \pi a^3 \ \rho_s, \\
&&\mathbf I_p = \frac{2 M_P}{5} \ a^2
\begin{pmatrix}
    1  & 0 & 0\\
    0  & 1 & 0\\
    0  & 0 & 1 
\end{pmatrix},
\end{eqnarray*}
where $\rho_s$ is  the particle density. For a porous particle of the porosity
$\tilde{P}$, we assume that the rigid skeleton is uniformly distributed over the entire volume of the porous ball so that its  mass and inertia tensor are
\begin{align*}
&M_P=\frac{4}{3} \pi a^3 (\tilde{P} \rho_f +(1-\tilde{P})\rho_s),\\
&\mathbf I_P =\frac{2M_p}{5} a^2
\begin{pmatrix}
    1  & 0 & 0\\
    0  & 1 & 0\\
    0  & 0 & 1 
\end{pmatrix},
\end{align*}
where $\rho_f$ is  the fluid density.  But for the neutrally buoyant ball considered in this article, the mass and inertia are the same as those of the rigid one since $\rho_s=\rho_f$.
\end{remark}

\subsubsection{Numerical results}\label{sec3.1.2}

\begin{table}
\centering
\caption{The rotational speed of the  porous ball of radius $a=0.1$ obtained by  $h=\frac{1}{48}$ and $CRIT=10^{-5}$} \label{table:1}
\begin{tabular}{||c c c||} 
 \hline
 $\Delta t$ & Permeability & $\omega(z)$  \\ [0.5ex] 
 \hline\hline
0.001 & 0.05 & -0.4998402 \\ 
0.001 & 0.01 & -0.4999398 \\
0.001 & 0.005 & -0.4998876 \\ 
0.001 & 0.0025 & -0.4997966 \\
0.001 & 0.001 & -0.4996004  \\
0.001 & 0.0005 & -0.4994060  \\
0.0005 & 0.00025 & -0.4992049  \\ [1ex] 
 \hline
\end{tabular}
\end{table}
\begin{table}
\centering
\caption{The rotational speed of the  porous ball of radius $a=0.15$ obtained by  $h=\frac{1}{48}$ and $CRIT=10^{-5}$}\label{table:2}
\begin{tabular}{||c c c||} 
 \hline
 $\Delta t$ & Permeability & $\omega$ (z)  \\ [0.5ex] 
 \hline\hline
0.001 & 0.05 & -0.4998634 \\ 
0.001 & 0.01 & -0.4995757 \\
0.001 & 0.005 & -0.4992370 \\ 
0.001 & 0.0025 & -0.4987220 \\
0.001 & 0.001 & -0.4978223  \\
0.001 & 0.0005 & -0.4971082  \\
0.0005 & 0.00025 & -0.4964840  \\ [1ex] 
 \hline
\end{tabular}
\end{table}
\begin{table}
\centering
\caption{The rotational speed of the  porous ball of radius $a=0.2$ obtained by  $h=\frac{1}{48}$ and $CRIT=10^{-5}$}\label{table:3}
\begin{tabular}{||c c c||} 
 \hline
 $\Delta t$ & Permeability & $\omega$ (z)  \\ [0.5ex] 
 \hline\hline
0.001 & 0.05 & -0.4996081 \\ 
0.001 & 0.01 & -0.4983921 \\
0.001 & 0.005 & -0.4972581 \\ 
0.001 & 0.0025 & -0.4957328 \\
0.001 & 0.001 & -0.4934660  \\
0.001 & 0.0005 & -0.4919084  \\
0.0005 & 0.00025 & -0.4906672  \\ [1ex] 
 \hline
\end{tabular}
\end{table}
\begin{figure}[ht!]
\begin{center}
(a)\includegraphics[width=.45\textwidth]{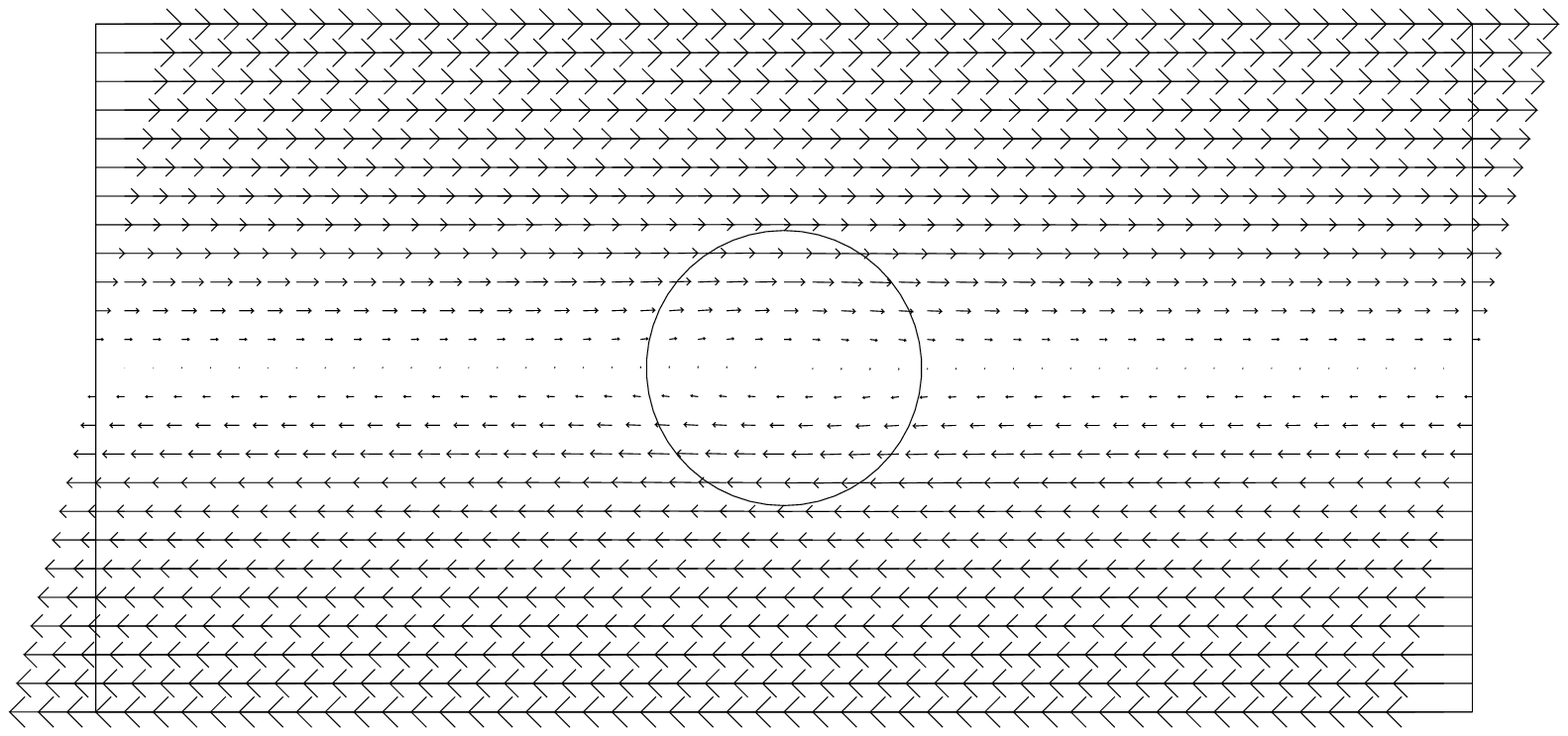}\  \includegraphics[width=.48\textwidth]{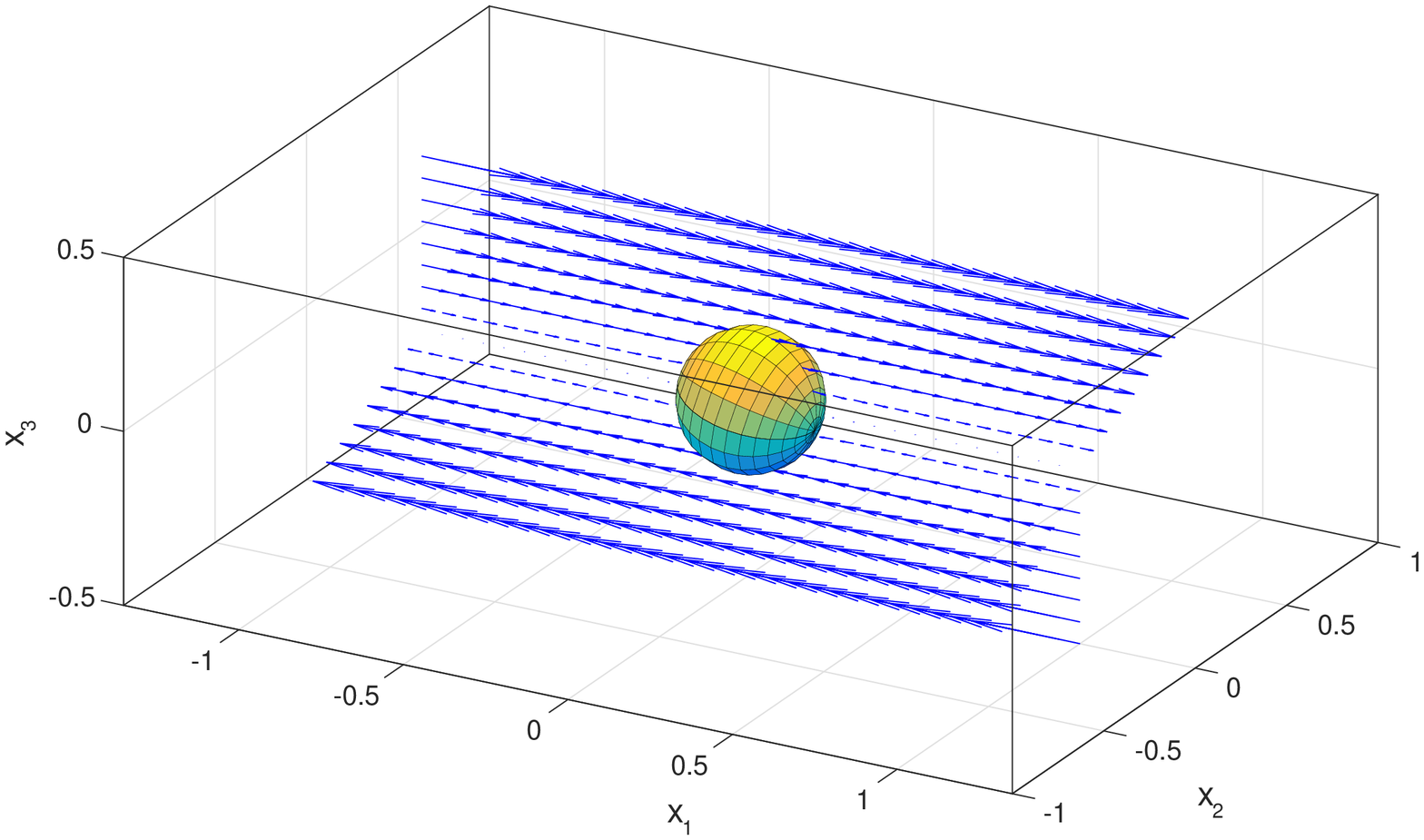}\\
(b)\includegraphics[width=.45\textwidth]{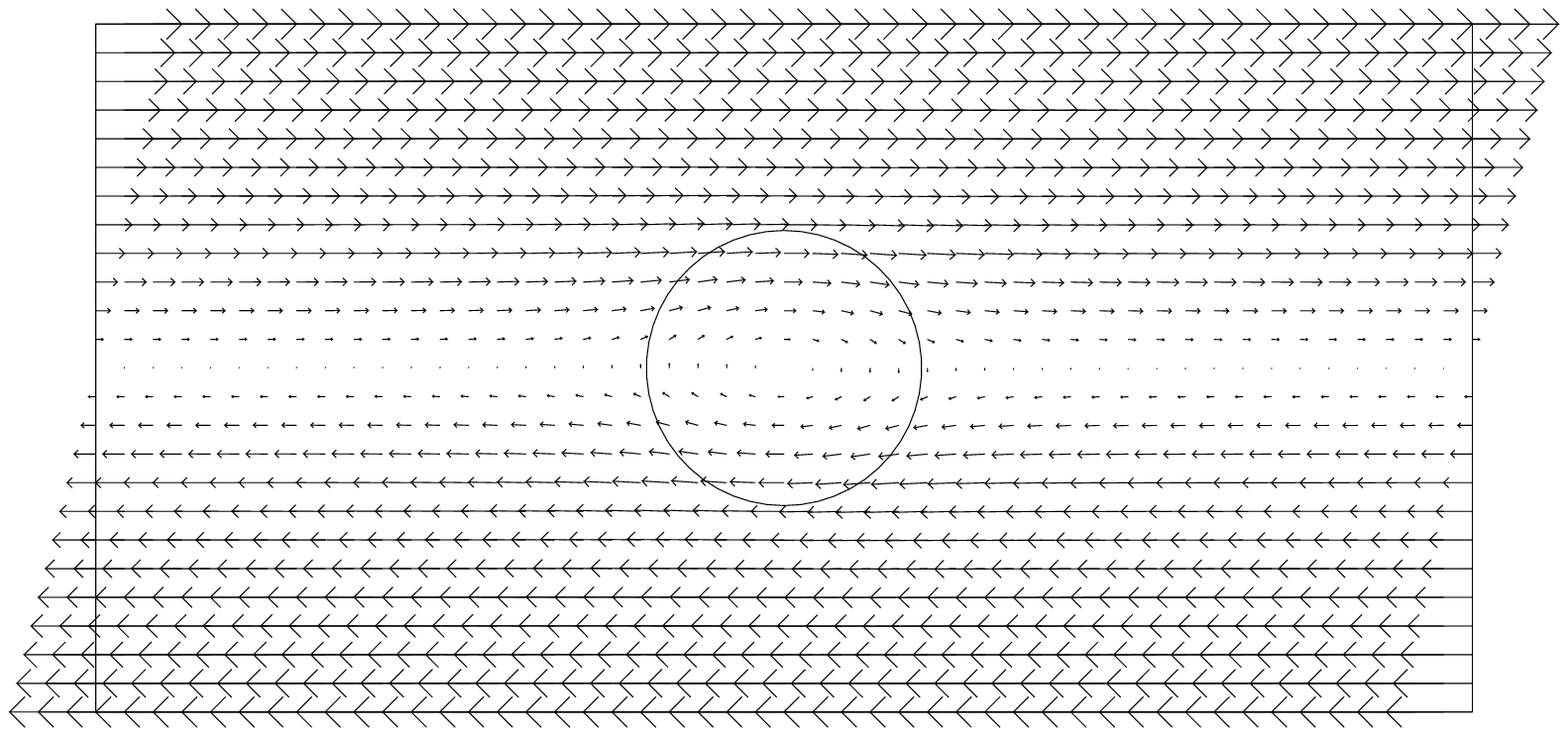}\  \includegraphics[width=.48\textwidth]{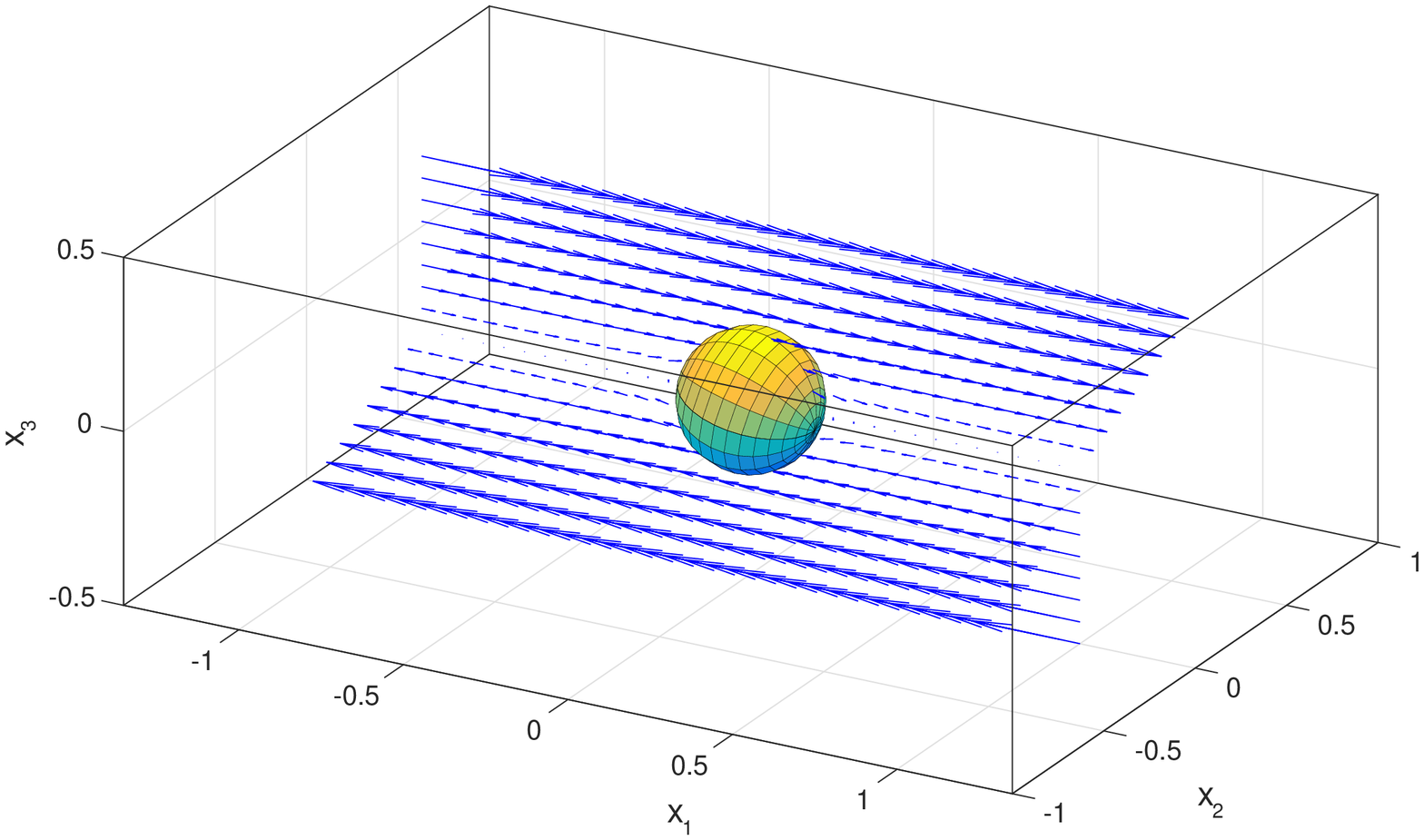}\\
(c)\includegraphics[width=.45\textwidth]{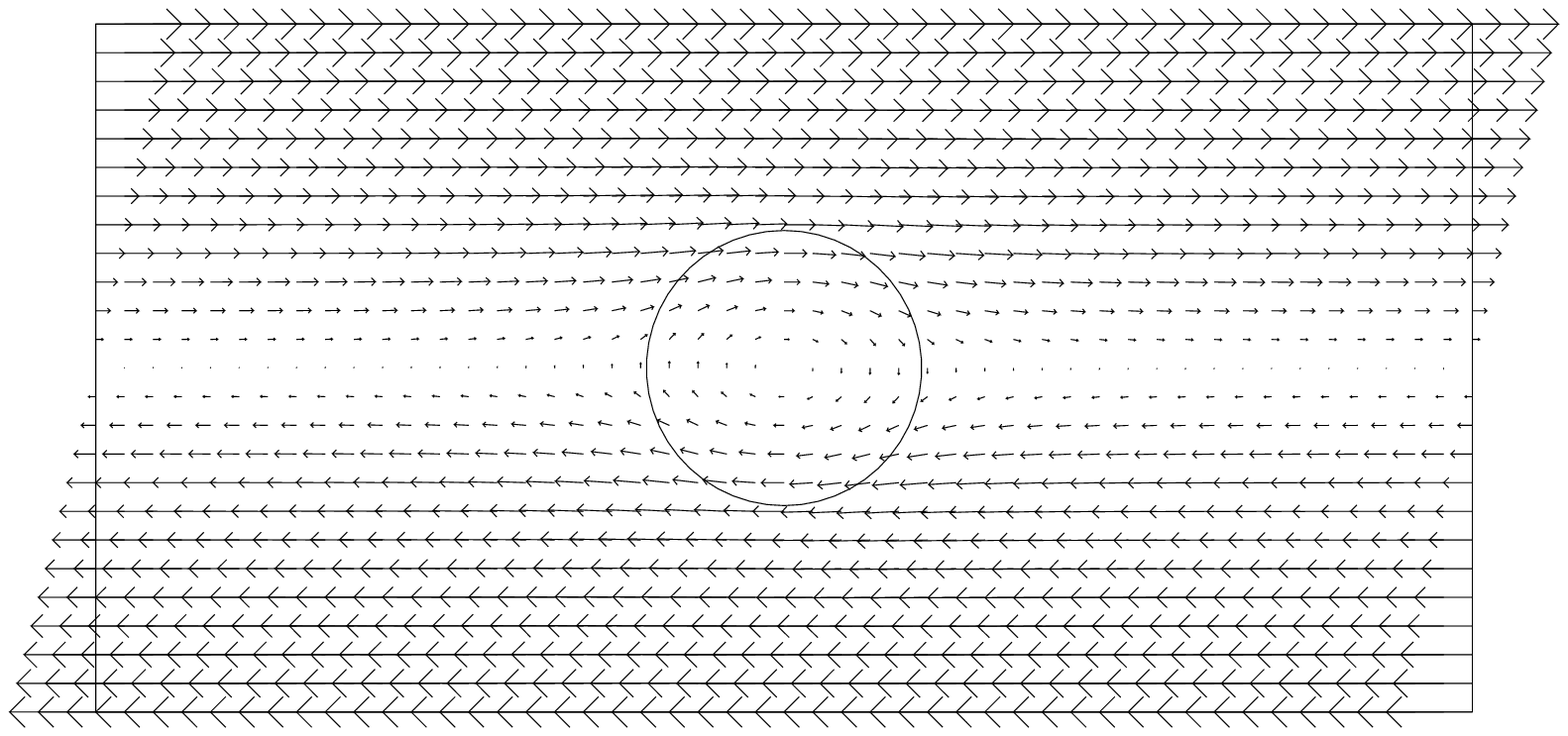}\ \includegraphics[width=.48\textwidth]{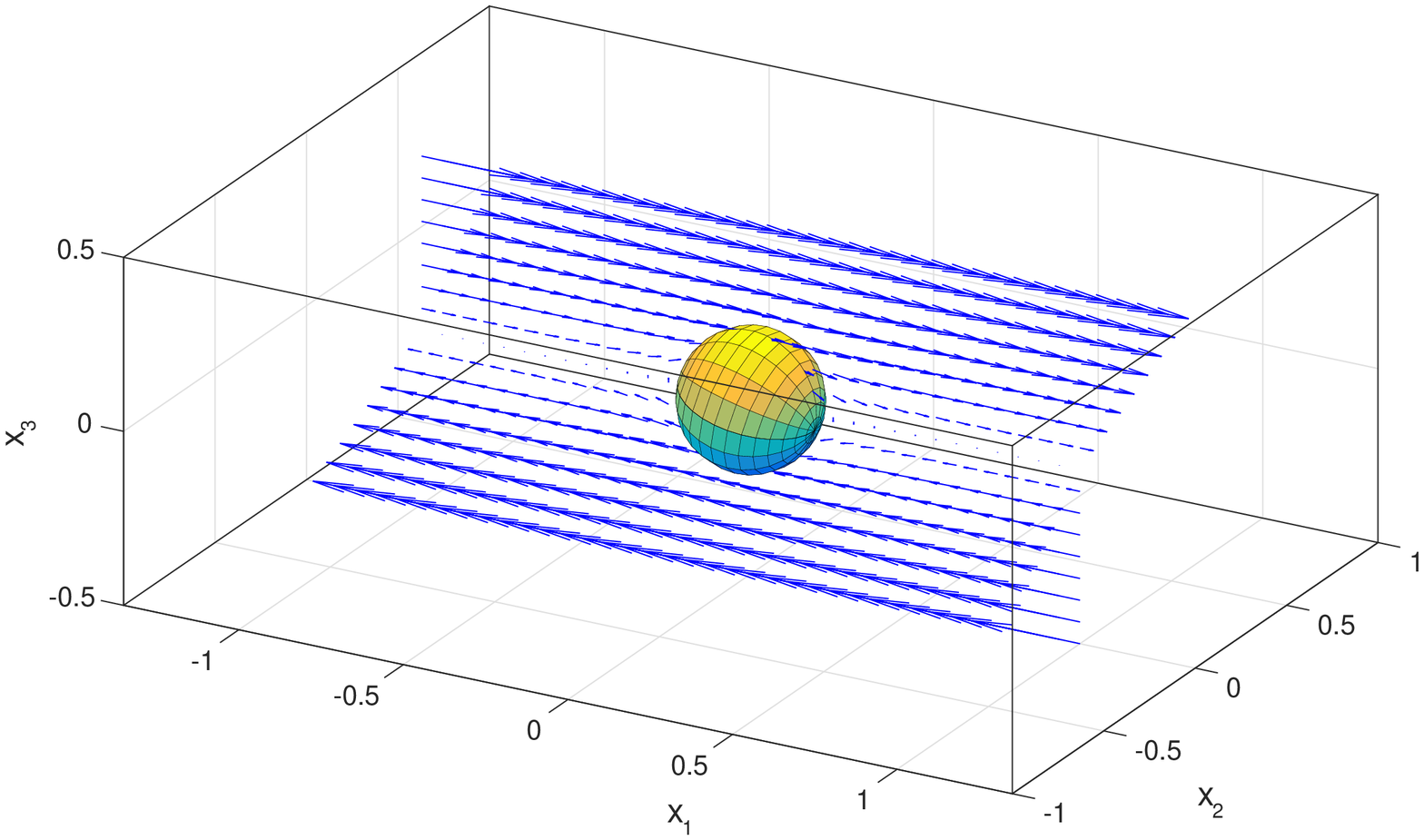}
\end{center}
\caption{Snapshots of the velocity field projected on the shear plane ($x_1x_3$-plane) for a porous 
ball of radius $0.2$ rotating in a bounded shear flow: the values of the permeability are (a) $k=$0.05,   (b) $k=$0.01, and 
   (c) $k=$0.005, respectively (from top to bottom). } \label{fig.2}
\end{figure}
\begin{figure}
\begin{center}
(d)\includegraphics[width=.45\textwidth]{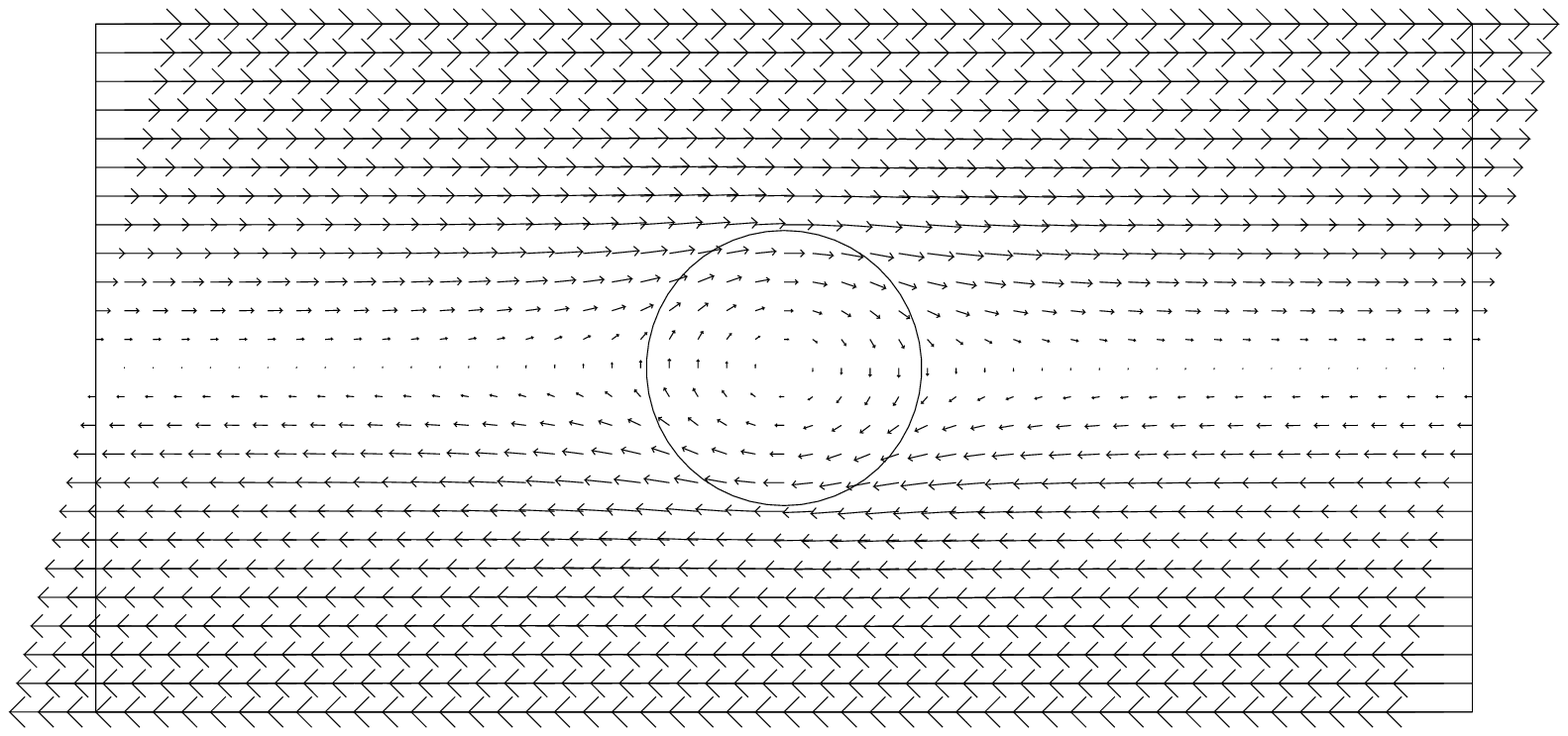}\ \includegraphics[width=.48\textwidth]{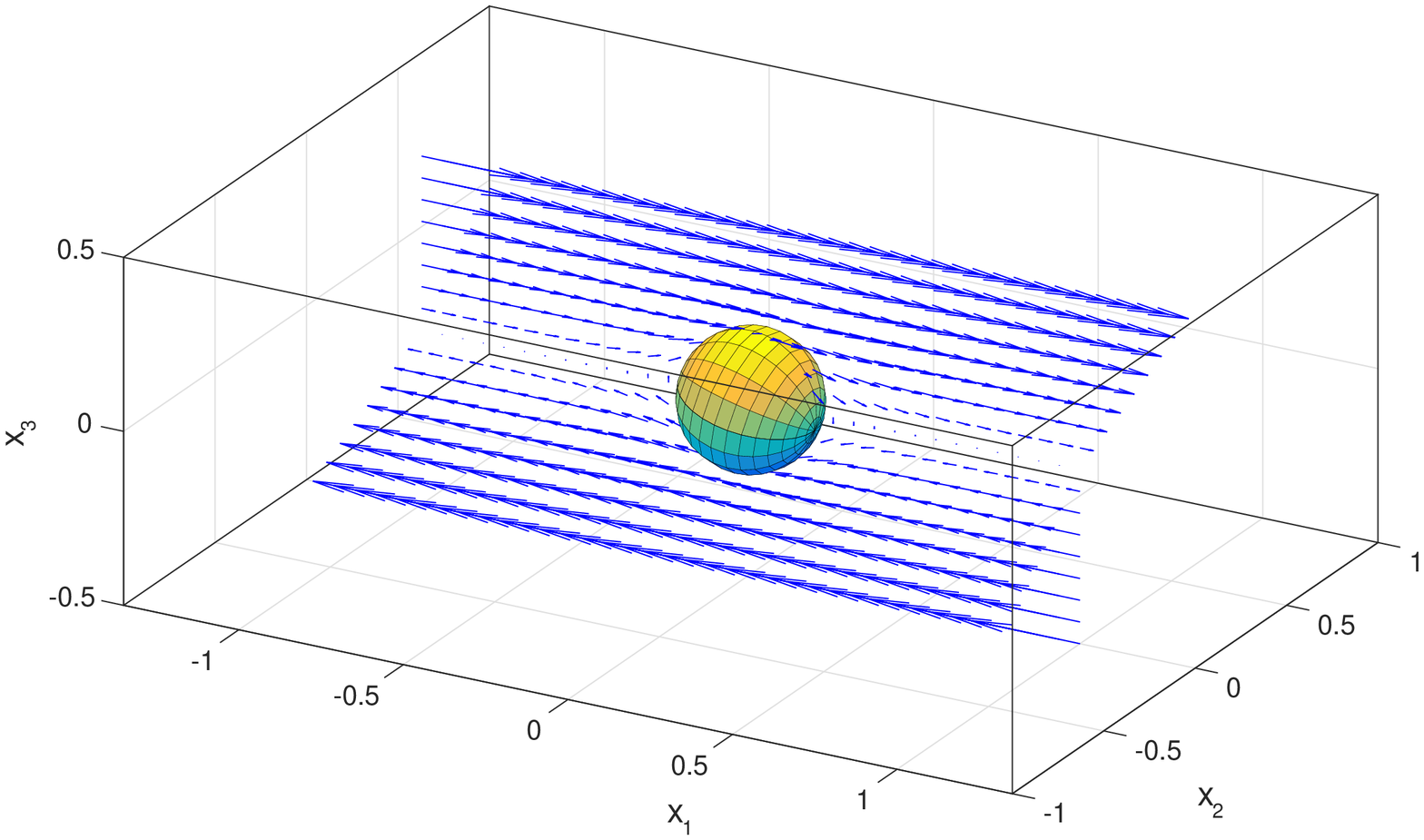}
(e)\includegraphics[width=.45\textwidth]{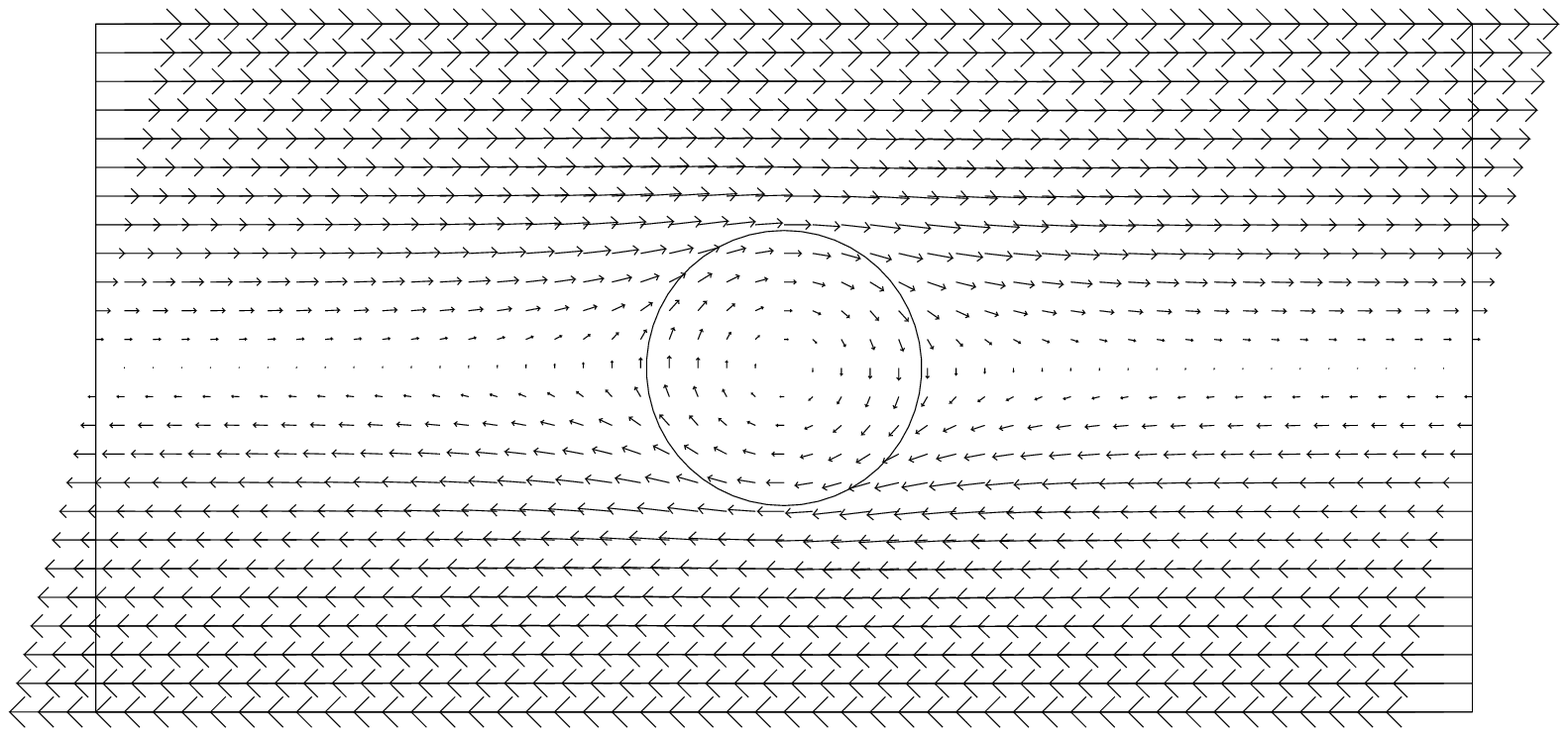}\   \includegraphics[width=.48\textwidth]{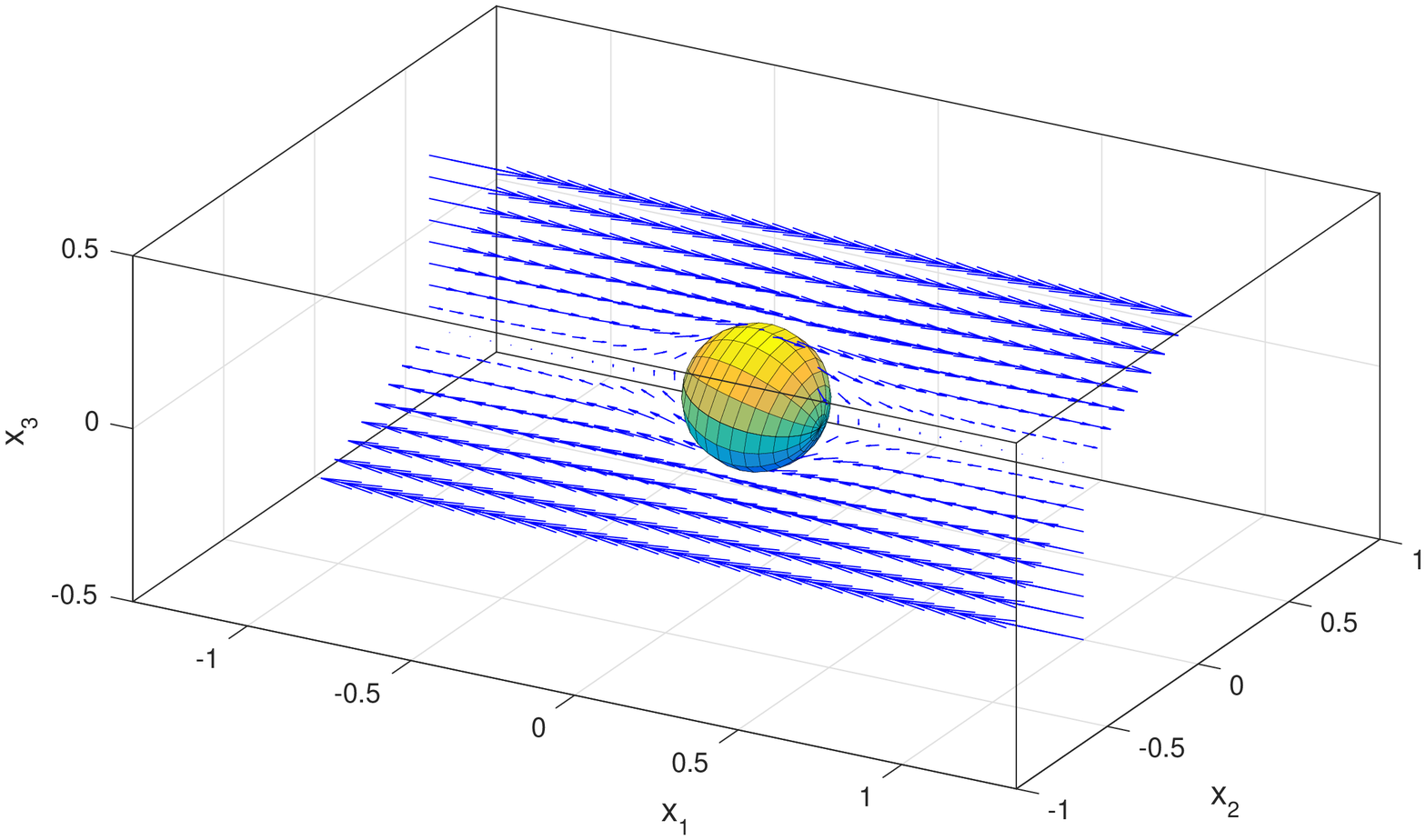}\\
(f)\includegraphics[width=.45\textwidth]{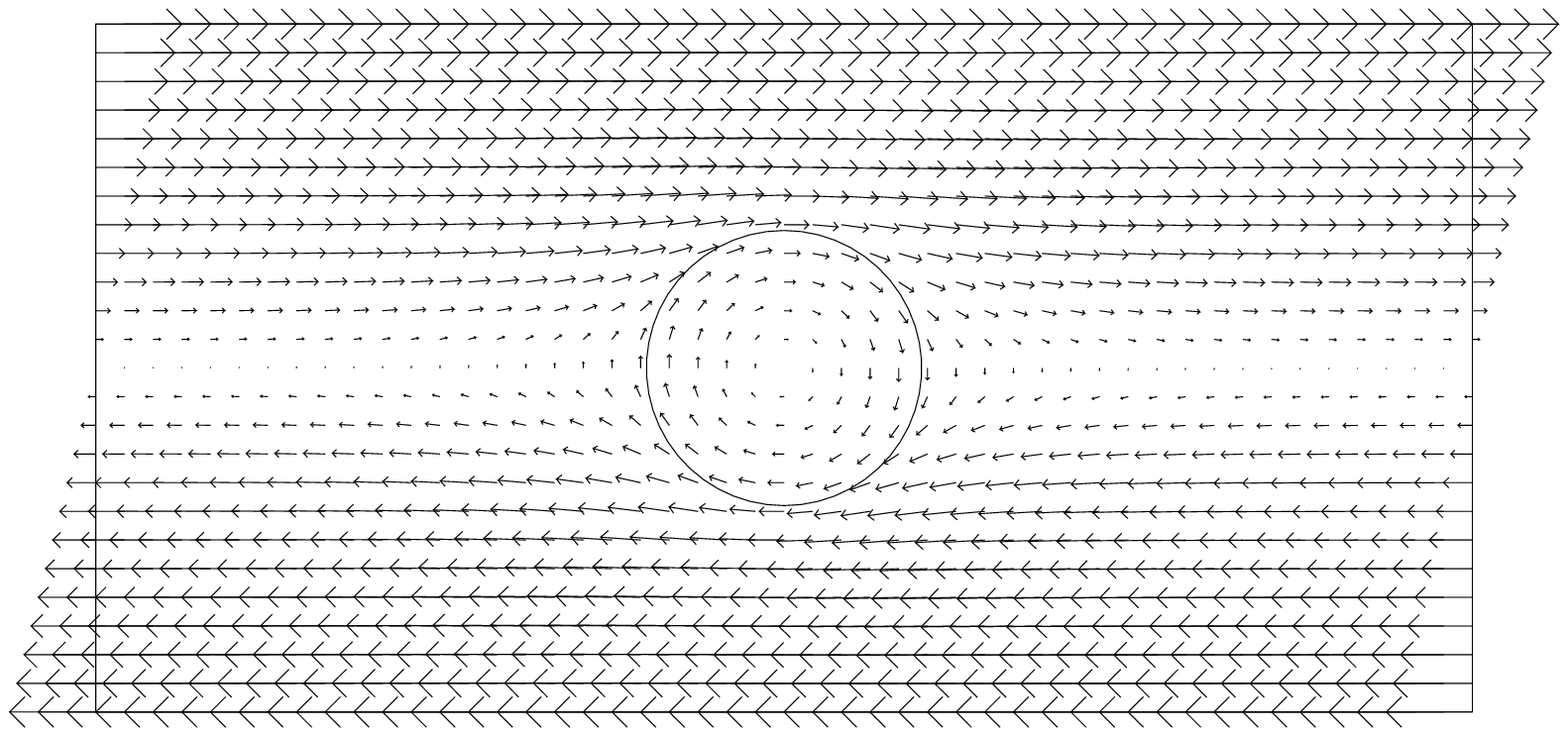}\  \includegraphics[width=.48\textwidth]{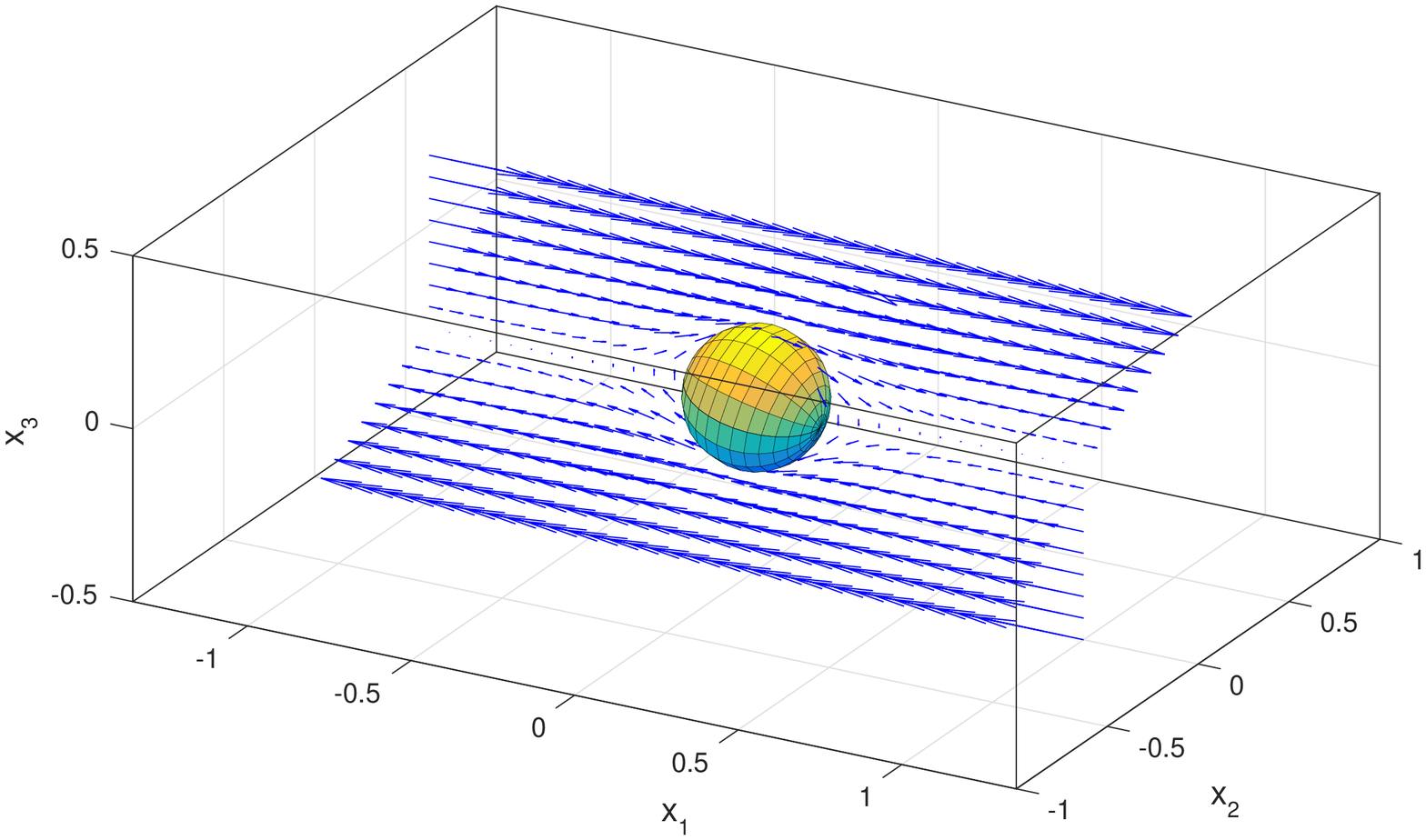}\\
(g)\includegraphics[width=.45\textwidth]{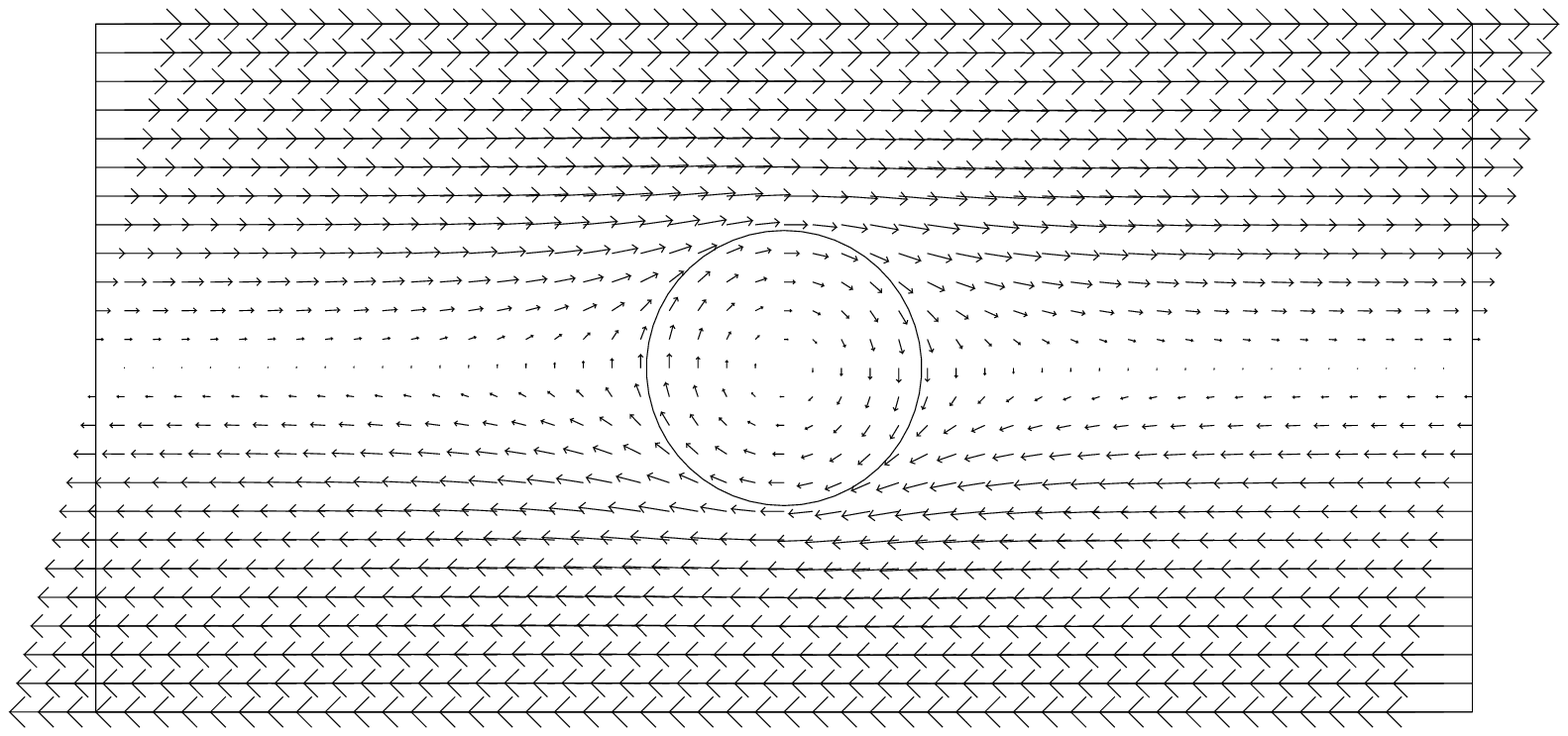}\ \includegraphics[width=.48\textwidth]{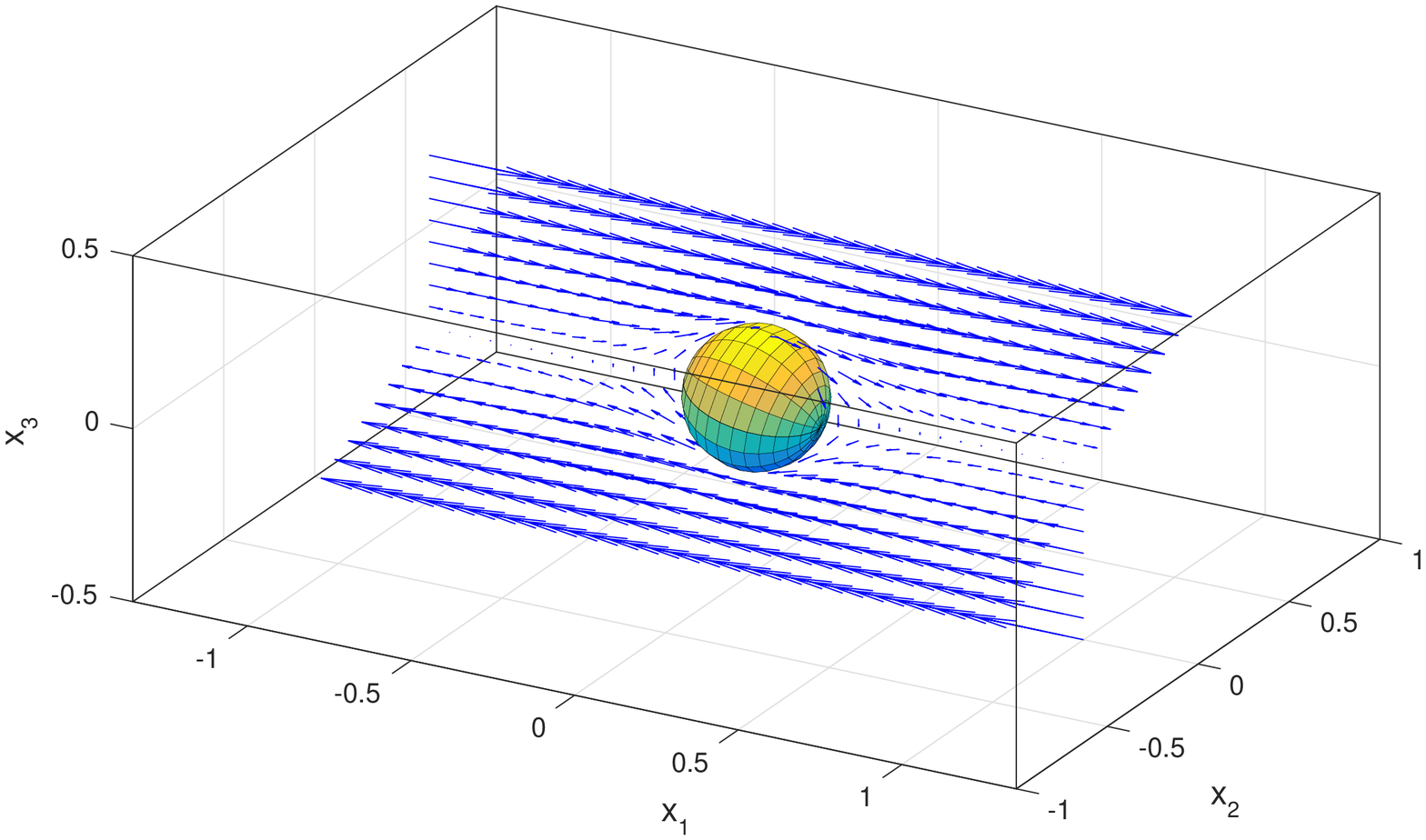}
\end{center}
\caption{Snapshots of the velocity field projected on the shear plane ($x_1x_3$-plane) for a porous 
ball of radius $0.2$ rotating in a bounded shear flow: the values of the permeability are   (d) $k=$0.0025,
   (e) $k=$0.001, (f) $k=$0.0005 and (g) $k=$0.00025, respectively (from top to bottom). } \label{fig.3}
\end{figure}
We have considered the cases of one porous ball freely rotating and suspended  initially at the middle between two moving walls as in Fig. \ref{fig.1}.  The computational domain is 
$\Omega = (-1, 1) \times (-1/2, 1/2) \times(-1, 1)$. The radius of the ball is $a$=0.1,
0.15, or 0.2 and the initial position of the ball mass center is at $(0,0,0)$.  The densities of the ball and fluid 
are both 1 and the fluid viscosity is also 1, thus the ball is neutrally buoyant.  The values of the permeability $k$ of the porous ball 
considered here are 0.05, 0.01, 0.005, 0.0025, 0.001, 0.0005 and 0.00025. The shear rate for the shear flow is  $\dot{\gamma}= 1$.
The space mesh size for the velocity field is $h =\frac{1}{48}$,
the time step being $\Delta t = 10^{-3}$.  The rotating speed of a rigid ball in an
unbounded shear flow is $-\dot{\gamma}/2$ according to the Jeffery's  solution \cite{Jeffery1922}. 
For different values of the permeability and the radius of the ball, the rotating speed of a porous rigid ball 
in a bounded shear flow is in a good agreement with the Jeffery's  solution as shown in Tables \ref{table:1},  
\ref{table:2} and \ref{table:3}.  The same conclusion has been obtained for the rotating speed
of a porous ellipsoid in shear flows reported in \cite{29}. But even if the rotating velocity inside the particle  is the same as the rigid body motion, 
the   fluid flow inside the porous ball does depend on the permeability. In Figs. \ref{fig.2} and \ref{fig.3}, the projection of the velocity field on the 
$x_1x_3$-plane clearly indicates  that the fluid motion is close to the rigid body motion only for much smaller value of
the permeability (e.g., $k=$0.0005 and 0.00025). For the higher value of  the permeability (e.g., $k=$0.05 and 0.01), the fluid just goes through the particle directly.

\subsection{Motion of two porous balls}

In this section we will consider numerical simulation of the interaction of two porous balls in a bounded shear flow. Due to that fact that  these two porous balls 
can move freely between the two moving walls, we will employ (\ref{eqn:14})-(\ref{eqn:18}) for simulating the dynamics of the two porous ball interaction.

\subsubsection{Description of the second numerical method}

The space-time discretization of (\ref{eqn:14})-(\ref{eqn:18}) is presented in the following 
(again, after dropping some of subscripts $h$):

\vskip 4ex
{\it For $n \ge 0$, $\mathbf{u^n,  V_i^n, G_i^n}$ and $\boldsymbol{\omega_i^n}$, $i=1$, $2$, being known (then  $B^n_{i}$, $i=1, 2$, are known),  
we solve the following sub-problems:}
\vskip 2ex
{\it Step 1. We compute  $\mathbf{u^{n+1}}$, $p^{n+1}$, $\mathbf V_i^{n+1}$  and $\boldsymbol \omega_i^{n+1}$, $i=1$, $2$,  via the solution of}
\begin{eqnarray}
&&\hskip -15pt \nu \int_\Omega \nabla \mathbf u^{n+1}: \nabla \mathbf v\, d \mathbf x - \int_\Omega p^{n+1} (\nabla \cdot \mathbf v) \, d \mathbf x \nonumber \\
&& \hskip 100pt =-\frac{\nu}{k} \sum_{i=1}^2 \int_{B^n_i}(\mathbf u^{n+1} - \mathbf u^{n+1}_{p,i})\cdot \mathbf v\, d \mathbf x, \ \forall \mathbf v \in \mathbf W_{0,h},\label{eqn:32}\\ 
&&\hskip -15pt \int_{\Omega} q \nabla \cdot  \mathbf u^{n+1} \, d \mathbf x =0, \ \forall q \in L^2_h,\label{eqn:33}\\
&&\hskip -15pt  M_{p,i} \frac{\mathbf V_i^{n+1}-\mathbf V_i^n}{\Delta t}=\frac{\nu}{k} \int_{B^n_i}(\mathbf u^{n+1}-\mathbf u^{n+1}_{p,i})d\mathbf x, \ i=1,2,\label{eqn:34}\\
&&\hskip -15pt  \mathbf I_{p,i} \frac{\boldsymbol \omega_i ^{n+1}-\boldsymbol \omega_i^n}{\Delta t} =\frac{\nu}{k} \int_{B^n_i} 
\overrightarrow{\mathbf G_i^n\mathbf x} \times (\mathbf u^{n+1}-\mathbf u^{n+1}_{p,i})\, d \mathbf x,\ i=1,2, \label{eqn:35}\\
&&\hskip -15pt  \mathbf u^{n+1} \in \mathbf W_h, \mathbf u^{n+1}={\mathbf g}_0(t^{n+1}) \textit{ on } \Gamma, \ p^{n+1} \in L^2_{0,h}.\label{eqn:36}
\end{eqnarray}
{\it Step 2. Update the particle mass centers:}
\begin{equation}
\mathbf G_i^{n+1}=\mathbf G_i^n +\mathbf V_i^{n+1} \Delta t,\ i=1,2. \label{eqn:37}
\end{equation}
In  (\ref{eqn:32})-(\ref{eqn:36}),\ we have  $\mathbf u^{n+1}_{p,i}(\mathbf x) = \mathbf V_i^{n+1} + 
\boldsymbol \omega_i^{n+1} \times \overrightarrow{\mathbf G_i^{n} \mathbf x}$,  $\forall \mathbf x \in B^n_i$ for $i=1, 2$. 

To solve problem (\ref{eqn:32})-(\ref{eqn:36}), we consider it as a discrete analog of the  following coupled system steady state:

{\it For a.e. $t >0$, find $\mathbf u(t) \in (H^1(\Omega))^3, \mathbf u= {\mathbf g}_0$ on $\Gamma$, $p(t)\in L^2_0(\Omega)$, 
$\mathbf V_i$, $\boldsymbol{\omega}_i \in \mathbb R^3$, $i=1, 2$, such that}
\begin{eqnarray}
&&\int_\Omega \frac{\partial \mathbf u}{\partial t} \cdot \mathbf v\, d \mathbf x + \nu \int_\Omega \nabla \mathbf u : \nabla \mathbf v\, d \mathbf x 
- \int_\Omega p \nabla \cdot \mathbf v dx \nonumber \\
&& \hskip 100pt =-\frac{\nu}{k} \sum_{i=1}^2 \int_{B_i}(\mathbf u - \mathbf u_{p,i})\cdot \mathbf v\, d \mathbf x, 
\forall \mathbf v \in \mathbf W_{0},\label{eqn:38} \\
&&\int_{\Omega} q (\nabla \cdot \mathbf \mathbf \mathbf u) \, d \mathbf x =0, \forall q \in L^2(\Omega),\label{eqn:39}\\
&&\mathbf u(0)=\mathbf u_0 \text{ (with } \nabla \cdot \mathbf u_0=0) \text{ in } \Omega,\label{eqn:40} \\
&&M_{p,i}(\mathbf V_i -\mathbf V_{0,i})=\frac{\nu \Delta t}{k} \int_{B_i}(\mathbf u-\mathbf u_{p,i}) \, d \mathbf x, \ i=1,2,\label{eqn:41}\\
&&\mathbf I_{p,i} (\boldsymbol \omega_i -\boldsymbol \omega_{0,i})=
\frac{\nu \Delta t}{k} \int_{B_i} \overrightarrow{\mathbf G_i^n\mathbf x} \times (\mathbf u-\mathbf u_{p,i})\, d \mathbf x, \ i=1,2.\label{eqn:42}
\end{eqnarray}
In system (\ref{eqn:38})-(\ref{eqn:42}), we have $\mathbf u_{p,i}(\mathbf x) = \mathbf V_i + \boldsymbol \omega_i \times \overrightarrow{\mathbf G_i^{n}\mathbf x}$,  
$\forall \mathbf x \in B_i$ for $i=1, 2$, where the mass center $\mathbf G_i^n$ is fixed, $\mathbf V_{0,i}$ and $\boldsymbol{\omega}_{0,i}$ are given for $i=1, 2$. 
Thus the region $B^n$ occupied by ball is fixed.

Combining the finite element approximations used in Section \ref{sec3.1} with he backward Euler scheme we obtain the following discrete analogue 
of (\ref{eqn:38})-(\ref{eqn:42}): 
\vskip 2ex
$\mathbf u^{(0)}=\mathbf u^n$, $\mathbf V^{(0)}_i= \mathbf V_i^n$,  
$\boldsymbol{\omega}_i^{(0)}=\boldsymbol{\omega}_i^n$,
$\mathbf u_{p,i}^{(0)}=\mathbf V_{i}^{(0)}+ \boldsymbol{\omega}_i^{(0)} \times \overrightarrow{\mathbf G_i^n \mathbf x}$,  $i=1, 2$, {\it are known}. 

{\it For  $m \ge 0$, $\mathbf{u^{(m)}}$, ${\mathbf V}_i^{(m)}$, $\boldsymbol{\omega}_i^{(m)}$, $\mathbf u^{(m)}_{p,i}$, $i=1, 2$  being known,  we solve the following 
sub-problems:}

{\it Step 1. We compute  $\mathbf u^{(m+1)} \textit{ and } p^{(m+1)}$ via the solution of}

\begin{eqnarray}
&&\int_{\Omega} \frac{\mathbf u^{(m+1)}-\mathbf u^{(m)}}{\tau}\cdot \mathbf v\, d \mathbf x +\nu \int_\Omega \nabla \mathbf u^{(m+1)}: \nabla \mathbf v\, d \mathbf x - \int_\Omega p^{(m+1)} (\nabla \cdot \mathbf v) \, d \mathbf x \nonumber \\
&& \hskip 100pt  = -\frac{\nu}{k} \sum_{i=1}^2 \int_{B_i}(\mathbf u^{(m)} - \mathbf u^{(m)}_{p,i})\cdot \mathbf v\, d \mathbf x, \quad \forall \mathbf v \in \mathbf W_{0,h},\label{eqn:43}\\ 
&&\int_{\Omega} q \nabla \cdot \mathbf \mathbf u^{(m+1)} \, d \mathbf x =0, \forall q \in L^2_h;\label{eqn:44}\\
&& \mathbf u^{(m+1)} \in \mathbf W_h, \mathbf u^{(m+1)}={\mathbf g}_0^{n+1} \textit{ on }  \Gamma, \ p^{(m+1)} \in L^2_{0,h}.\nonumber
\end{eqnarray}

{\it Step 2. Update the values}
\begin{eqnarray}
&& M_{p,i}  {\mathbf V_i^{(m+1)}-\mathbf V_{0,i}}=\frac{\nu {\Delta t}}{k} \int_{B_i}(\mathbf u^{(m+1)}-\mathbf u^{(m)}_{p,i}) \, d \mathbf x,\ i=1,2,\label{eqn:45}\\
&& \mathbf I_{p,i}  {\boldsymbol \omega_i ^{(m+1)}-\boldsymbol \omega_{0,i}} =\frac{\nu {\Delta t}}{k} \int_{B^n_i} \overrightarrow{G_i^n\mathbf x} \times (\mathbf u^{(m+1)}-\mathbf u^{(m)}_{p,i})\, d \mathbf x,\ i=1,2,\label{eqn:46}\\
&&\textit{and set} \nonumber\\
&& \mathbf u_{p,i}^{(m+1)}=\mathbf V_{i}^{(m+1)}+ \boldsymbol \omega_i^{(m+1)} \times \overrightarrow{\mathbf G_i^n \mathbf x}, \quad \forall \mathbf x \in B_i,\  i=1,2.\label{eqn:47}
\end{eqnarray}

\begin{figure}[t!]
\begin{center}
\includegraphics[width=0.4\textwidth]{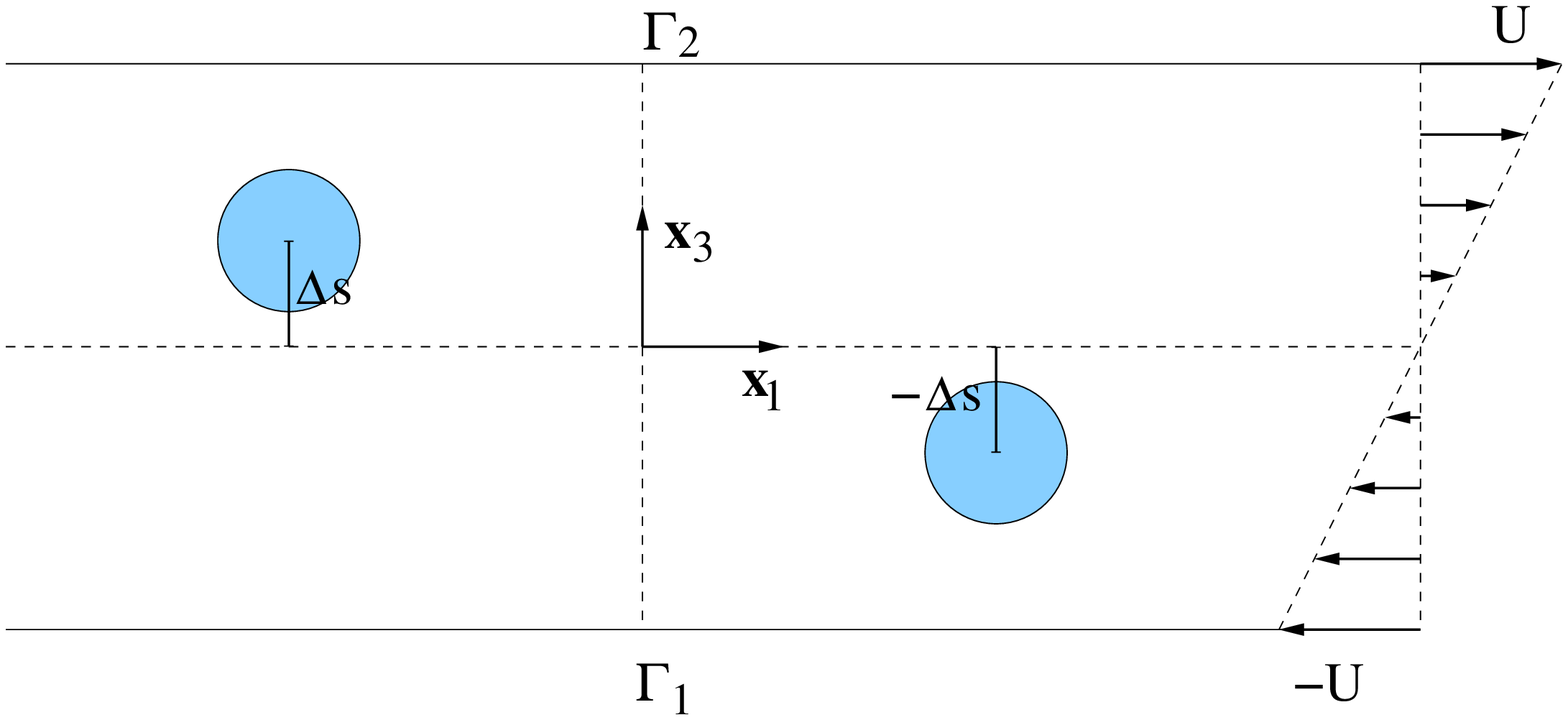} \ \ \
\includegraphics[width=0.5\textwidth]{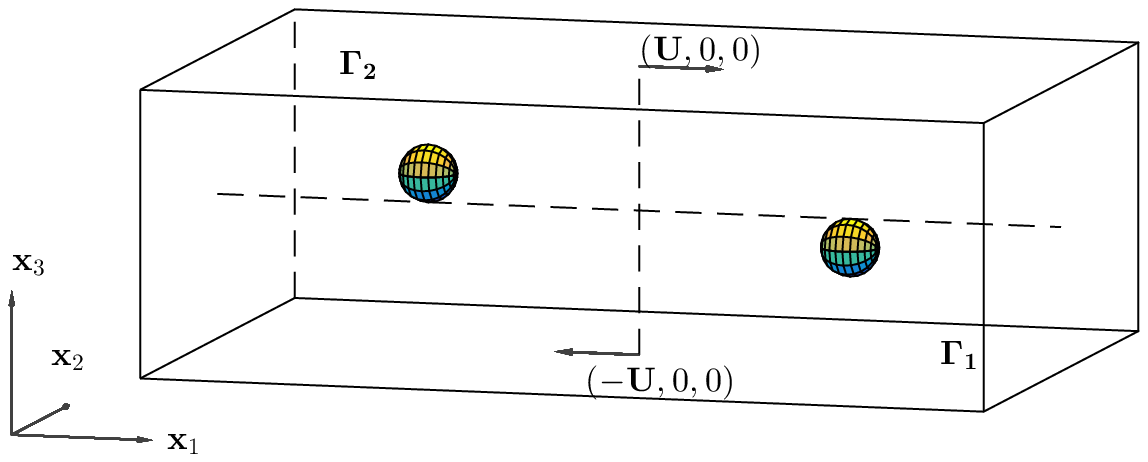}
\end{center}
\caption{The initial position of two porous balls in a bounded shear flow region.} \label{fig.4}
\end{figure}

\begin{remark} We have, again, used $\mathbf u^{(m)}$ in the right hand side of equation (\ref{eqn:43}) so that a fast solver can be applied 
at each iteration when solving system (\ref{eqn:43})-(\ref{eqn:47})  by an Uzawa/preconditioned
conjugate gradient algorithm    operating in the space $L^2_{0h}$ as discussed in, e.g., \cite{glowinski2003}. 
For the cases of two porous balls  suspended in a bounded shear flow, where we are looking for a steady state solution, 
we stoped algorithm (\ref{eqn:43})-(\ref{eqn:47}) as soon as 
\begin{equation*}
\frac{1}{\tau} \|\mathbf u^{(m+1)}-\mathbf u^{(m)}\| < CRIT,
\end{equation*}
where {\it CRIT} is the tolerance. The numerical results obtained in the Section \ref{sec3.2.2} have been obtaned with $CRIT=10^{-5}$.
\end{remark}

\begin{remark}
For the two ball interaction in a bounded shear flow,  there is no lubrication force between the two balls
under creeping  flow conditions.  Therefore, we are not allowed to apply an artificial repulsive
force to prevent ball overlapping in numerical simulation since such force might alter the trajectories of the two ball mass centers. 
To deal with the interaction during the two ball interaction, 
we have to impose a minimal gap of size $ch$ between the balls where $c$ is some constant between 0 and 1, $h$ being the 
mesh size of the velocity field. Then, when advancing the two ball mass centers in equation (\ref{eqn:37}), we proceed as follows at each sub-cycling time step: 
(i) we do nothing if the gap between the two balls at the new position is greater or equal than $ch$,
(ii) if the gap size of the two balls at the new position is less than $ch$, we do not advance the balls directly; but instead we first move the ball centers
in the direction perpendicularly to the line joining the previous centers, and then move them in the direction parallel to the line joining the previous centers, 
and make sure  that the gap size is no less than $ch$.  
\end{remark}

\begin{figure}[ht!]
       \centering
            \includegraphics[width=0.675\textwidth]{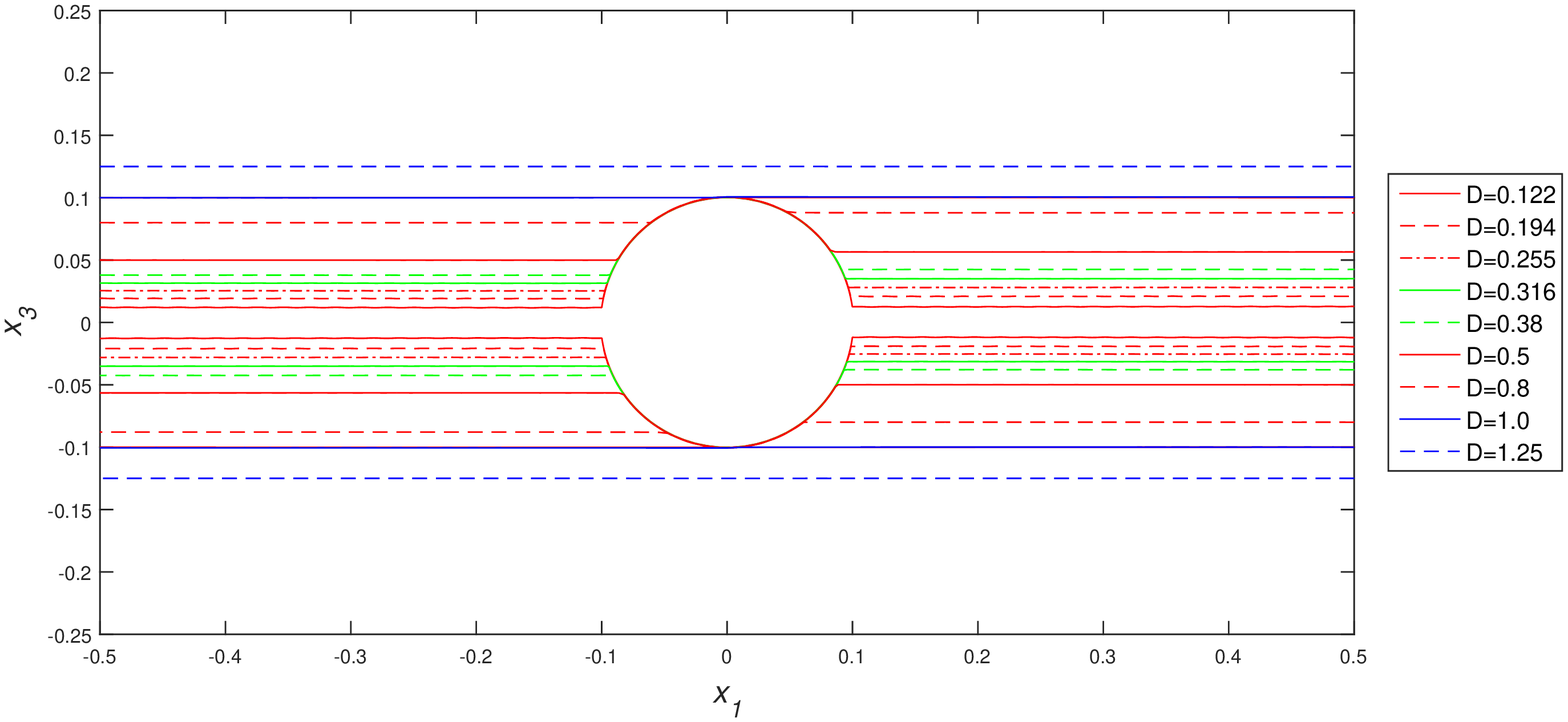}
            \includegraphics[width=0.675\textwidth]{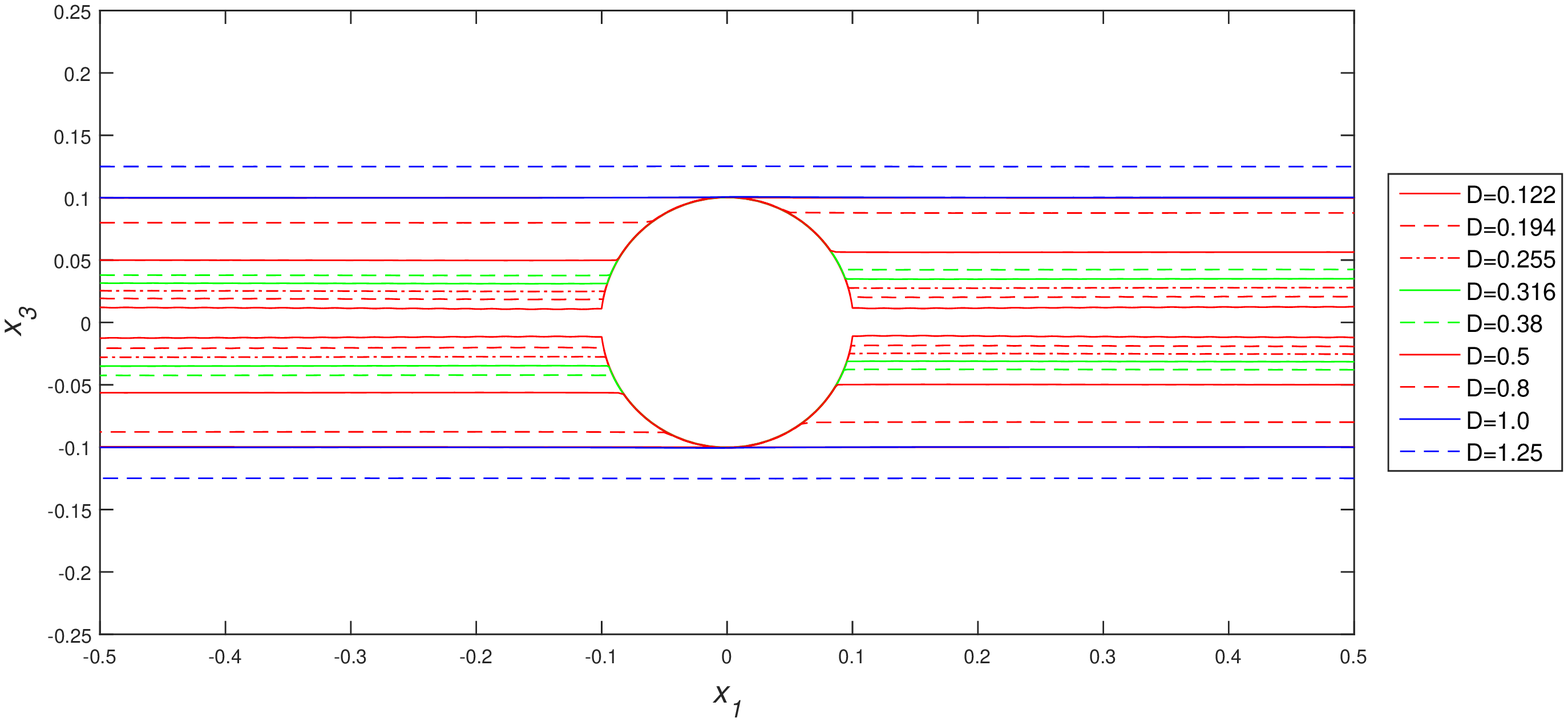}
            \includegraphics[width=0.675\textwidth]{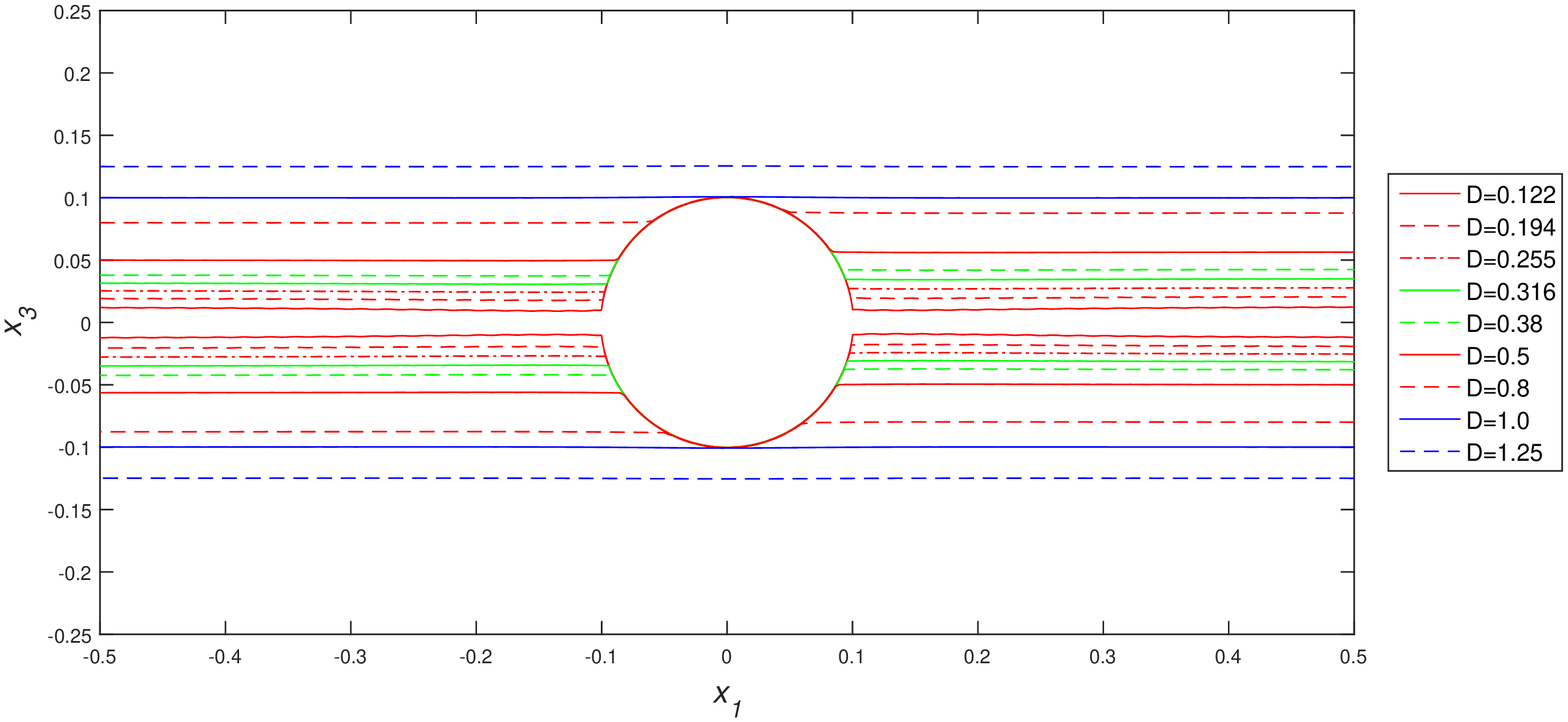}
            \includegraphics[width=0.675\textwidth]{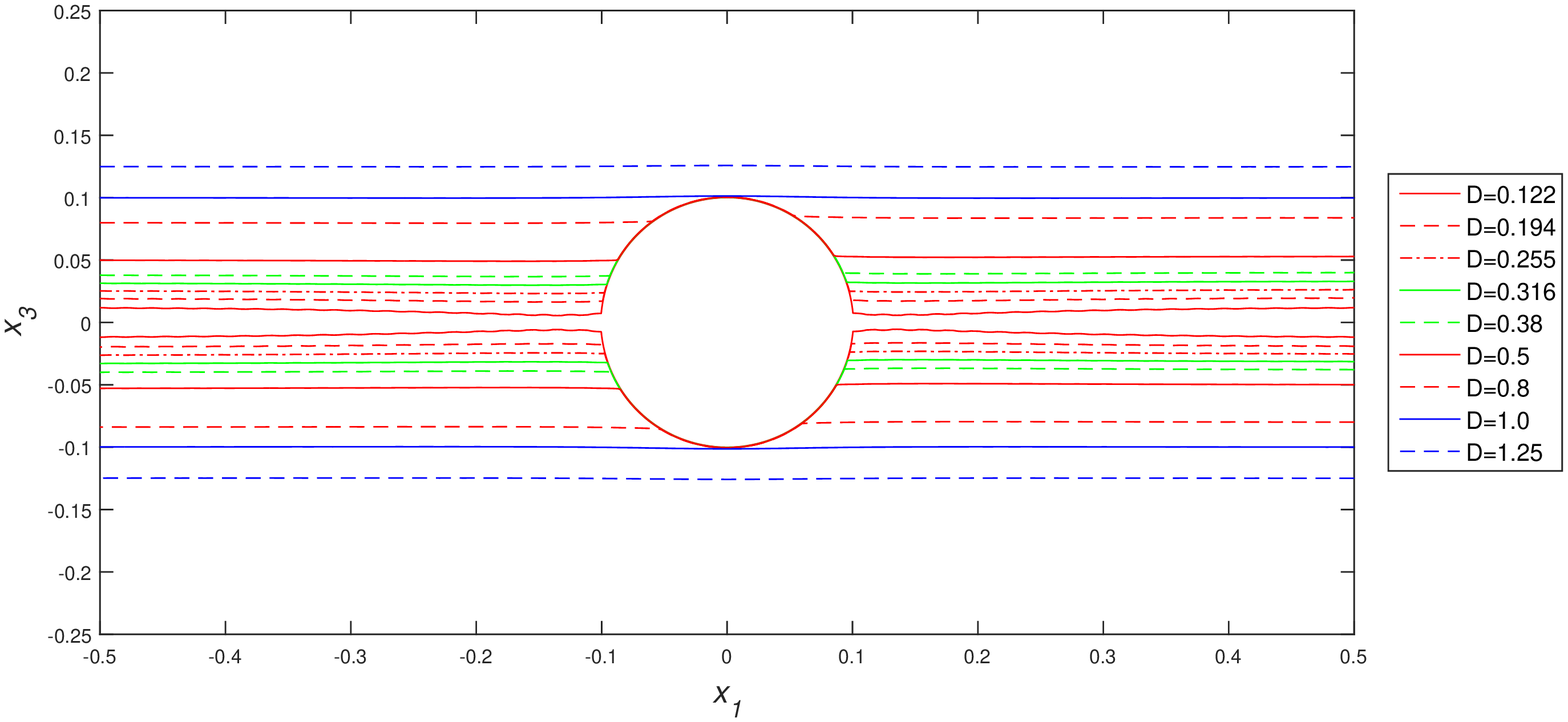}
            \caption{Trajectories of the ball mass centers in a bounded shear flow for the  permeability $k=$0.05, 0.01, 0.005 and 0.0025 
             from top to bottom).} \label{fig.5}
\end{figure}  
\begin{figure}[ht!]
       \centering
            \includegraphics[width=0.675\textwidth]{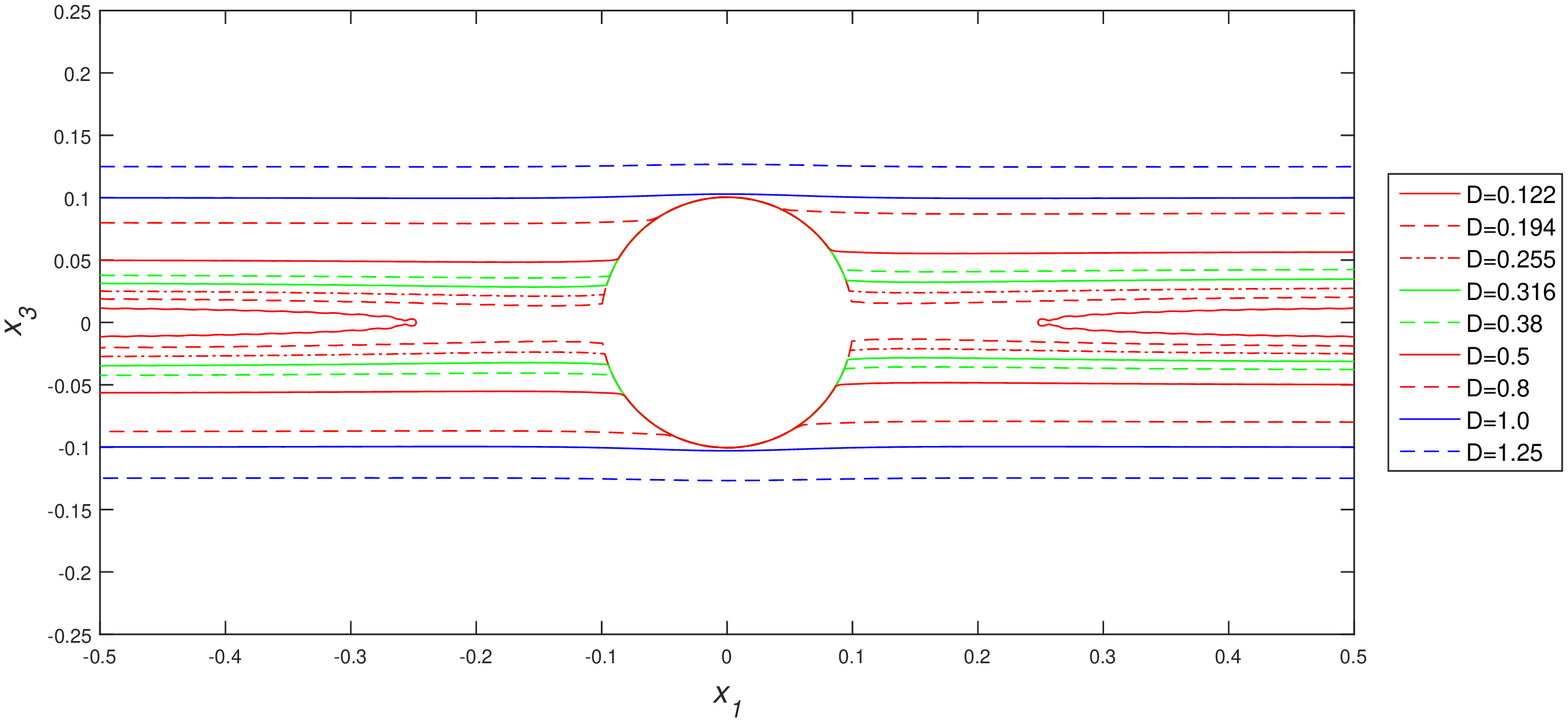}
            \includegraphics[width=0.675\textwidth]{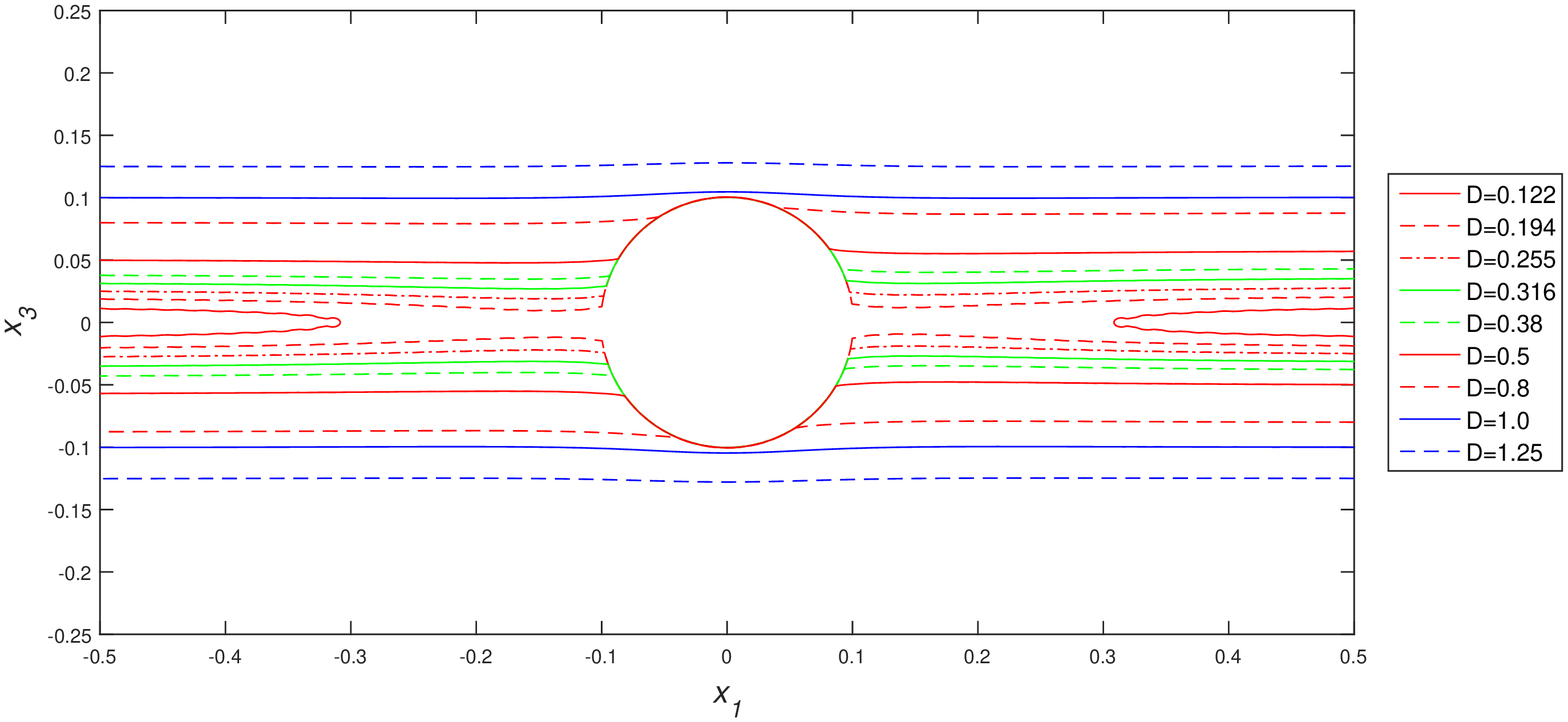}
            \includegraphics[width=0.675\textwidth]{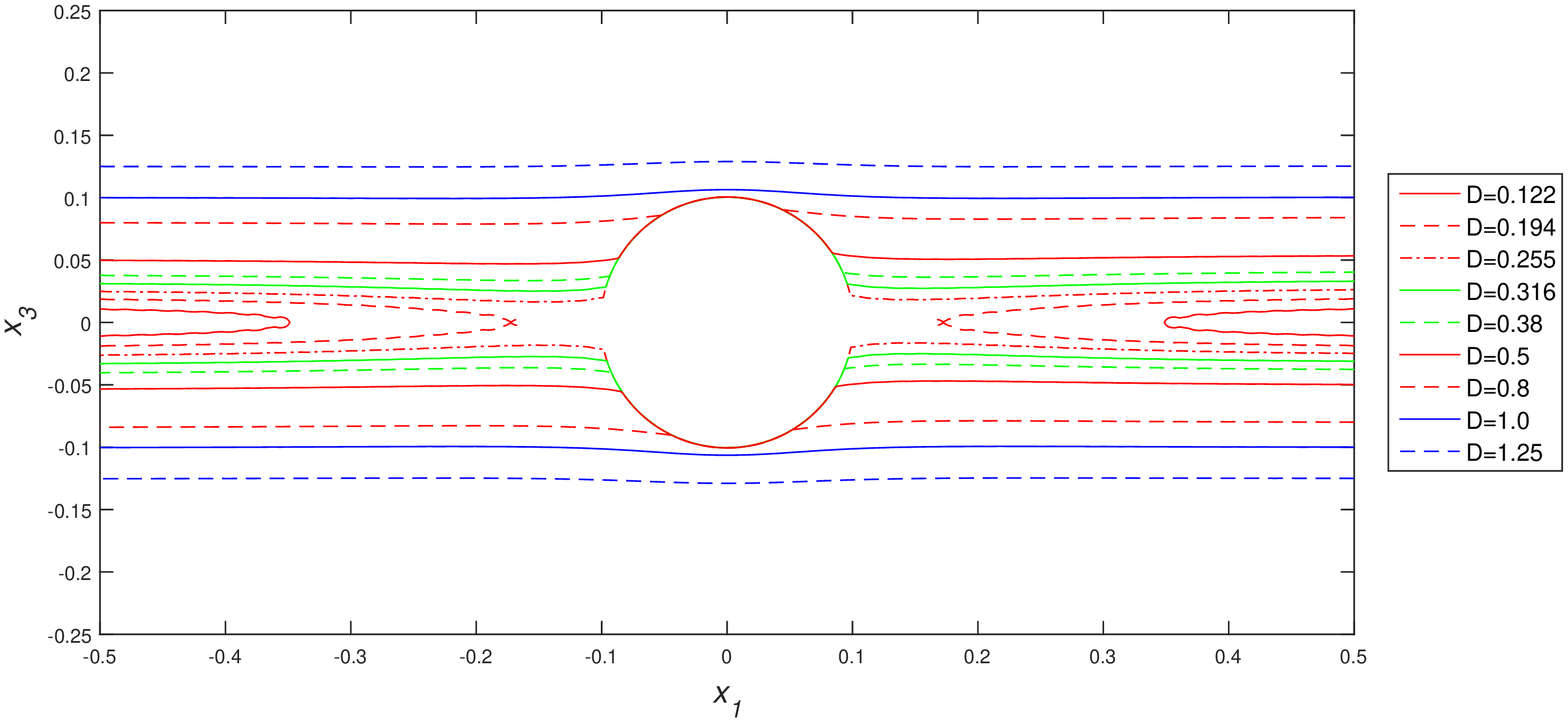}
            \caption{Trajectories of the ball mass centers in a bounded shear flow for the  permeability $k=$0.001, 0.0005 and 0.00025
             from top to bottom).} \label{fig.6}
\end{figure} 
\begin{figure}[pth!]
\centering
 \includegraphics[width=0.47\textwidth]{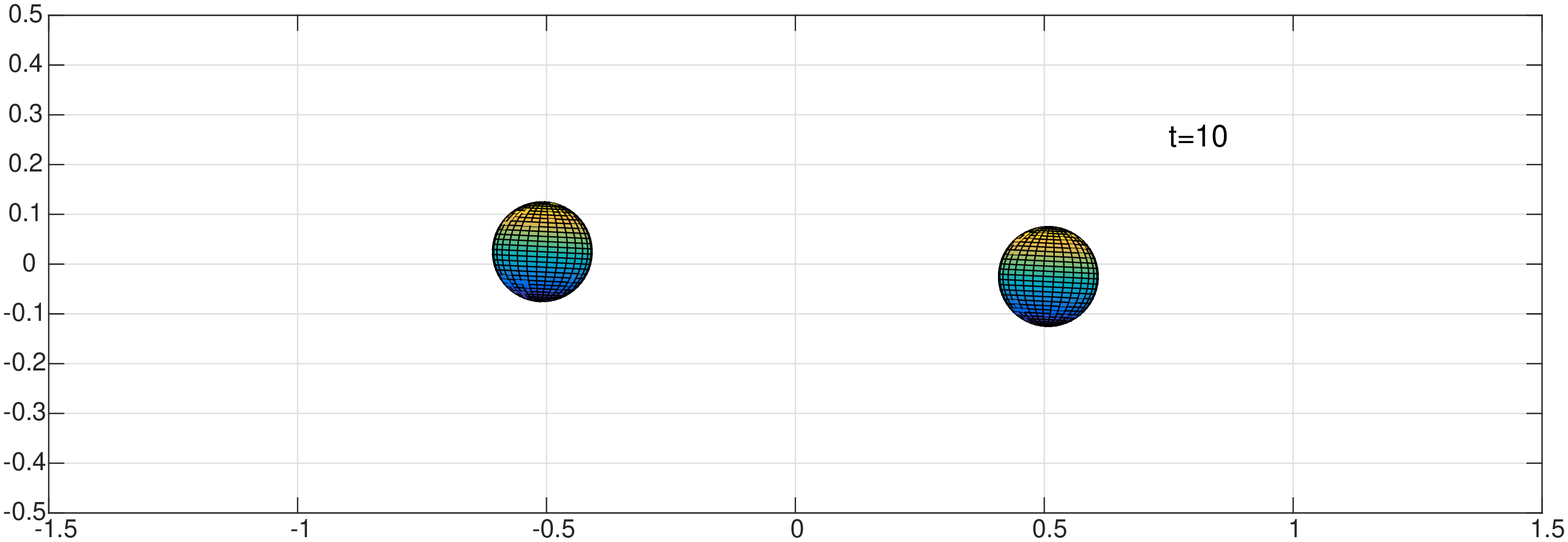} \\
 \includegraphics[width=0.47\textwidth]{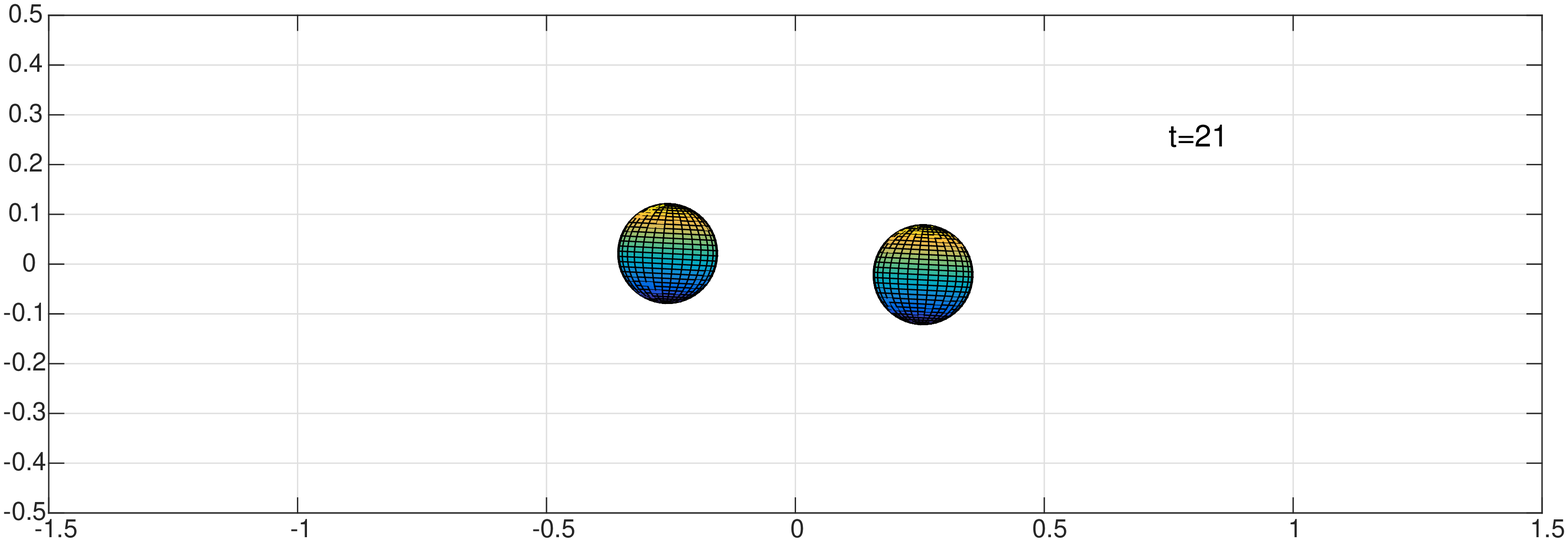} \\
 \includegraphics[width=0.47\textwidth]{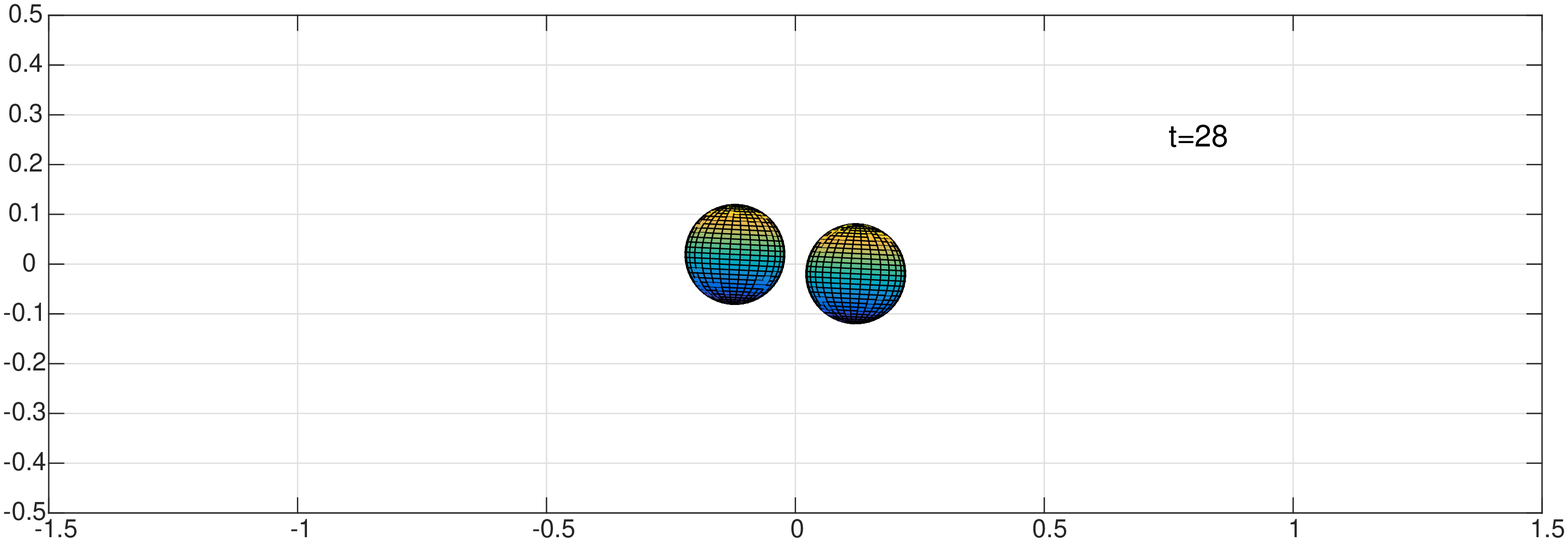} \\
 \includegraphics[width=0.47\textwidth]{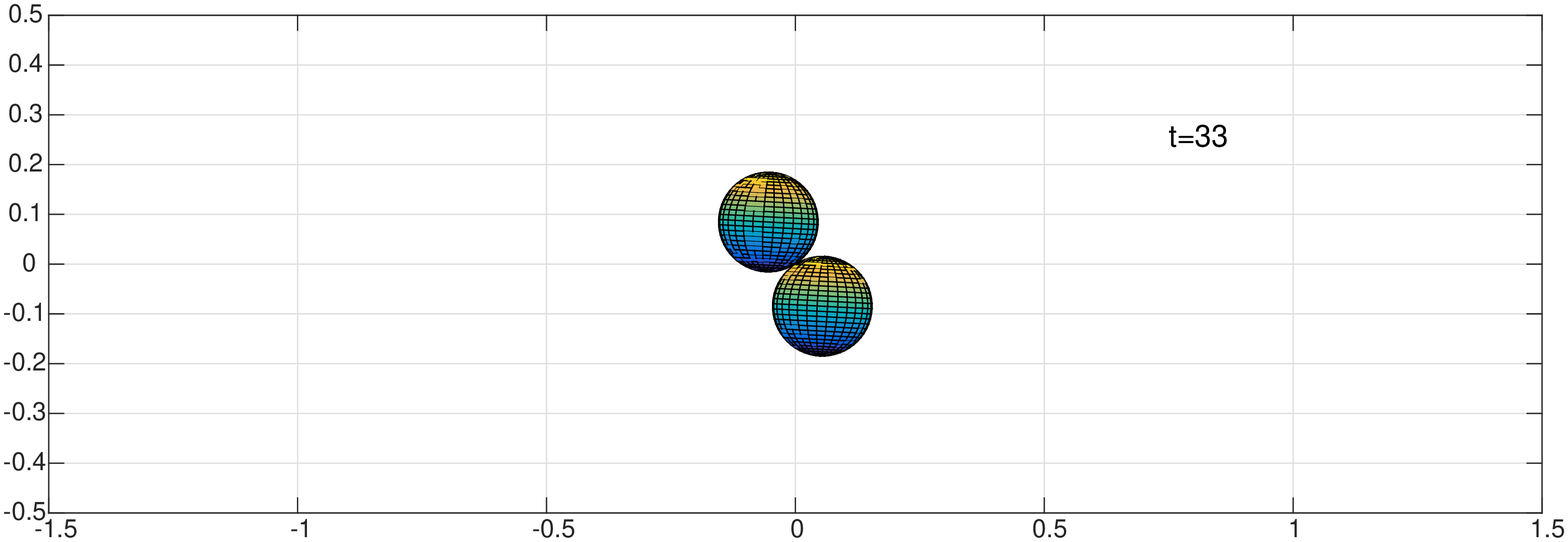} \\
 \includegraphics[width=0.47\textwidth]{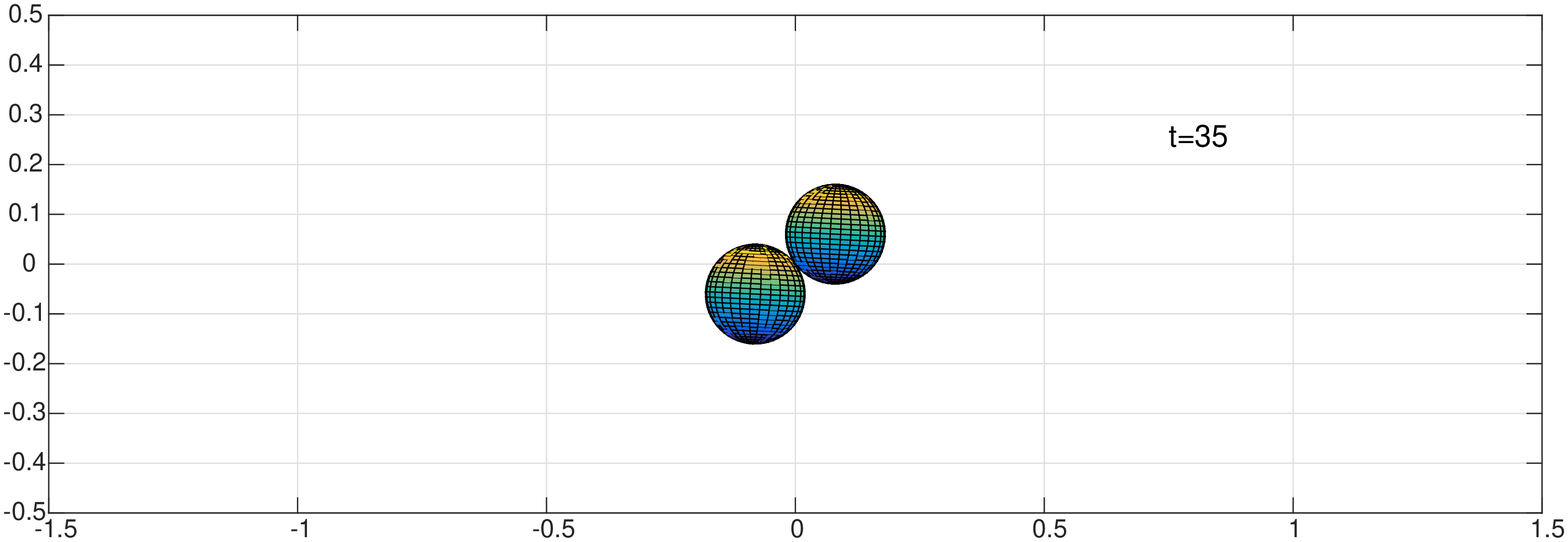} \\
 \includegraphics[width=0.47\textwidth]{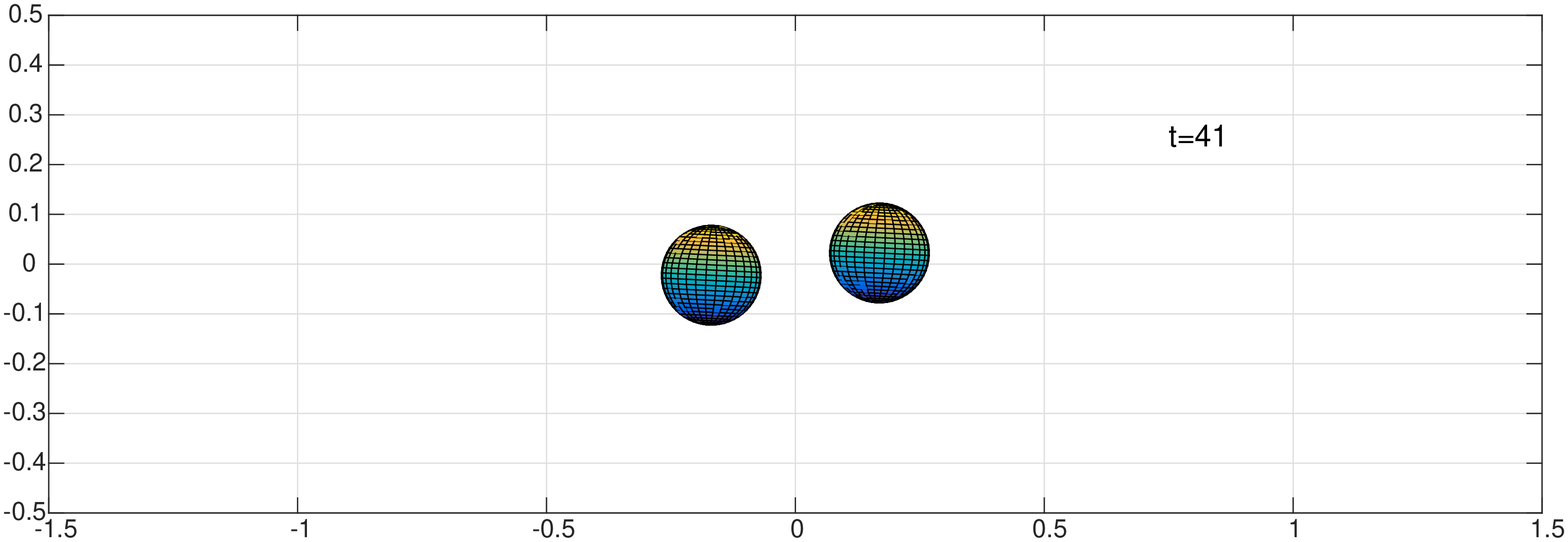} \\
 \includegraphics[width=0.47\textwidth]{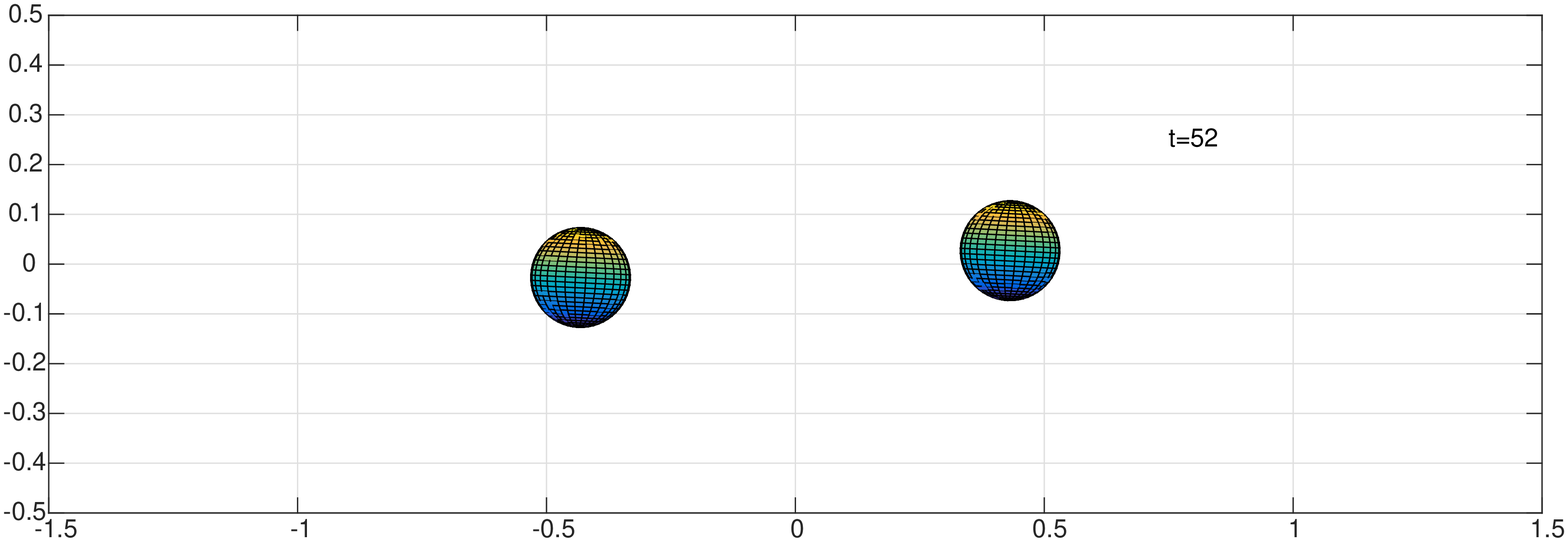}    
\caption{The snapshots of the particle position for the permeability $k=0.0005$ and $D=\Delta s/2a$=0.255 at $t=$10,   21, 
  28,   33, 35, 41, and  52  (from top to bottom and left to right).}\label{fig.7}
\end{figure}

\subsubsection{Numerical results}\label{sec3.2.2}

We have considered the cases of two porous balls freely suspended initially on the $x_1x_3$-plane as in Fig. \ref{fig.4}.  The computational domain is 
$\Omega = (-1.5, 1.5) \times (-1/2, 1/2) \times(-1, 1)$. The two ball radii are $a=$ 0.1.
The initial position of the two ball mass centers are at $(-0.75,0,\triangle s)$ and $(0.75,0,-\triangle s)$ for a given vertical
displacement $\triangle s > 0$ (see Fig. \ref{fig.4}). The values of the vertical displacement are $D=\Delta s/2a =$0.122, 0.194, 0.255, 0.316, 0.38, 0.5, 0.8, 1.0
and 1.255. The densities of the ball and fluid  are both 1 and the fluid viscosity is also 1. Thus the balls are neutrally buoyant.  
The values of the permeability $k$ of the porous ball are 0.05, 0.01, 0.005, 0.0025, 0.001, 0.0005 and 0.00025. 
The shear rate for the shear flow is 1. The space mesh size for the velocity field is $h =\frac{1}{48}$.
The time step is $\Delta t = 0.001$ for $k=$0.05, 0.01, 0.005, 0.0025, 0.001, and 0.0005 and  $\Delta t = 0.0005$ for $k=$0.00025.

In the Stokes regime, the interaction of the two non-porous and rigid balls in simple shear flows has been well studied (see, e.g., \cite{Zurita2007}).
Zurita {\it et al.} obtained that, depending of their vertical initial height $\triangle s$, the balls pass over/under each other or rotate around the midpoint 
of their mass centers if they are located as in Fig. \ref{fig.4} initially.
The first kind is called the non-swapping of the trajectories of the two ball mass centers, i.e., the higher one takes over the lower one and then both 
return to their initial heights while moving in a bounded shear flow. The second kind is   called the swapping of the trajectories, i.e., 
they come close to each other and to the mid-plane between the two horizontal walls, then, the balls move away from each other and from the above mid-plane.
For those wondering about the influence of porosity on the ball interaction let us mention the following: For the higher values of the 
permeability (e.g., $k=$ 0.05, 0.01, 0.005 and 0.0025), the two porous balls pass over/under each other for the vertical  displacements 
as in Fig. \ref{fig.5}. Also for the cases of smaller values of $k$ in Fig. \ref{fig.6}, the two porous balls still have non-swapping trajectories for most
values of the vertical displacement. For these non-swapping  between two porous balls, the two balls first come toward to each other, then almost touch each other and 
rotate with respect to the midpoint between two ball mass centers, and finally separate (see Fig. \ref{fig.7} for the details).  Thus, the interaction of the two 
non-swapping porous balls is quite different from the one between two non-porous and rigid balls as reported in \cite{Zurita2007}.  When  viewing 
the vector field in Figs. \ref{fig.2} and \ref{fig.3},  it is not surprising to see that the streamlines go through the porous ball, which explain why we have obtained 
such kind of non-swapping  trajectories for the  two porous balls. But for the lower permeability cases, the  swapping between two porous balls are obtained for smaller 
values of the vertical displacement (e.g., $\Delta s/2a$=0.122 for $k=$ 0.001 and 0.005 and $\Delta s/2a$=0.122 and 0.184 for $k=$0.00025).

\section*{Acknowledgments.} 
 
We acknowledge  the support of NSF (grant DMS-1418308).

\end{document}